\documentclass[referee, pdflatex,sn-mathphys-num]{sn-jnl}
\usepackage{afterpage}
\usepackage{algorithm}
\usepackage{algorithmicx}
\usepackage{algpseudocode}
\usepackage{amsfonts}
\usepackage{amsmath}
\usepackage{amssymb}
\usepackage{amsthm}
\usepackage[title]{appendix}
\usepackage{booktabs}
\usepackage{etoolbox}  
\usepackage{float}     
\usepackage{graphicx}
\usepackage{hyperref}

\usepackage{listings}
\usepackage{lmodern}
\usepackage{manyfoot}
\usepackage{mathrsfs}
\usepackage{multirow}
\usepackage{pifont}    
\usepackage{placeins}
\usepackage{svg}
\usepackage{tabularx}
\usepackage{textcomp}
\usepackage{verbatim}
\usepackage{xcolor}

\DeclareMathAlphabet{\mathscr}{OMS}{cmsy}{m}{n}

\newcommand{\cmark}{\ding{51}} 
\newcommand{\xmark}{\ding{55}} 

\newcommand{\yes}{\cmark}
\newcommand{\no}{\xmark}

\makeatletter
\@ifundefined{@kludgeins}{%
  \newinsert\@kludgeins
  \global\dimen\@kludgeins\maxdimen
  \global\count\@kludgeins 1000
}{}
\@ifundefined{@makespecialcolbox}{%
  \gdef\@makespecialcolbox{...}%
}{}
\@ifundefined{@reinserts}{%
  \gdef\@reinserts{%
     \ifvoid\footins\else\insert\footins{\unvbox\footins}\fi
     \ifvbox\@kludgeins\insert\@kludgeins{\unvbox\@kludgeins}\fi
  }%
}{}
\makeatother

\lstdefinelanguage{Haskell}{
  morekeywords={data,type,where,newtype,deriving,module,import,let,in,if,then,else,case,of,do,
    class,instance,default,infix,infixl,infixr,forall,Shells,Nothing,Just,Orbital,Shell,
    Coordinate,SubShell,Molecule,Eq,Read,Show,Double,Integer,Maybe,ElementAttributes,BondType,
    Map,Atom,AtomicSymbol,FilePath,IO,Meas,Int,Bond,Reaction,Condition,TempCondition,
    PressureCondition,Times,Element,Angstrom,PositiveDouble},
  sensitive=true,
  morecomment=[l]{--},
  morecomment=[s]{\{-}{-\}},
  morestring=[b]"
}

\lstdefinelanguage{C++}{
  morekeywords={typedef,adjacency_list,graph_traits,vertex_iterator,vertices,remove_vertex,tie,
    listS,vecS,Graph,N,for,if,else,while,return,include,namespace,std,int,double,char,void,bool,
    public,private,protected,class,struct,template,typename,this,new,delete,const,virtual,friend,
    operator,inline,static,using},
  sensitive=true,
  morecomment=[l]{//},
  morecomment=[s]{/*}{*/},
  morestring=[b]"
}

\lstdefinestyle{HaskellStyle}{language=Haskell}
\lstdefinestyle{CppStyle}{language=C++}

\newtheoremstyle{thmstyleone}{}{}{}{}{}{}{\newline}{}
\theoremstyle{thmstyleone}

\newtheorem{example}{Example}

\newtheorem{definition}{Definition}
\begin{document}

\title[Representing Molecules with Algebraic Data Types: Beyond SMILES and SELFIES]{Representing Molecules with Algebraic Data Types: Beyond SMILES and SELFIES}

\author*[1]{\fnm{Oliver} \sur{Goldstein}}\email{oliverjgoldstein@gmail.com}
\author*[2]{\fnm{Samuel} \sur{March}}\email{sammarch2@gmail.com}

\affil[1]{\orgdiv{Department of Computer Science}, \orgname{University of Oxford},
  \orgaddress{\street{Wolfson Building, Parks Road}, \city{Oxford}, \postcode{OX1 3QD}, \country{United Kingdom}}}

\affil[2]{\orgname{Independent researcher}, \city{Bristol}, \country{United Kingdom}}



\abstract{

Benchmarks of molecular machine learning models often treat the molecular representation as a neutral input format, yet the representation defines the syntax of validity, edit operations, and invariances that models implicitly learn. We propose MolADT, a typed intermediate representation (IR) for molecules expressed as a family of algebraic data types that separates (i) constitution via Dietz-style bonding systems, (ii) 3D geometry and stereochemistry, and (iii) optional electronic annotations. By shifting from string edits to operations over structured values, MolADT makes representational assumptions explicit, supports deterministic validation and localized transformations, and provides hooks for symmetry-aware and Bayesian workflows. We provide a reference implementation in Haskell (open-source, archived with DOI) and worked examples demonstrating delocalised/multicentre bonding, validation invariants, reaction extensions, and group actions relevant to geometric learning. \medskip

\textbf{Scientific Contribution:} We (1) introduce a representation-level framework that treats molecular representations as well-defined syntactic contracts rather than serializations, (2) formalize a layered typed IR capturing constitution/geometry/annotations, and (3) provide an open reference implementation intended to enable more controlled and interpretable benchmarking of molecular ML pipelines.

}
\maketitle

\enlargethispage{\baselineskip} 

\section{Introduction}\label{intro}

The choice of molecular representation affects cheminformatics workflows such as molecular property prediction and generative design \citep{krenn2022selfies}. Different representations can enable the exploration of different classes of molecules, as well as relationships between a molecule's chemistry and it's properties.  Different representations can also make it easier to structure the information required by machine learning models. The mathematical primitive used in a representation, and  data type (i.e. the ``format'' of a representaion on a computer) can both constrain and illuminate workflows of a cheminformatician \citep{McBride2016Type}. 

Because representation choices can change what is easy to learn, what is considered ``valid'' and what constitutes a meaningful perturbation, model benchmarks are only comparable when the representational substrate is specified and stable. Our goal is not to report new state-of-the-art prediction metrics; rather, we propose a representation designed to make representational assumptions explicit and therefore benchmarkable. \medskip

We propose treating molecular representation as a typed intermediate representation: a structured value with explicit invariants and well-scoped edit operations, rather than as a surface syntax (SMILES/SELFIES) whose syntax are enforced only by external parsers/normalisers. Benchmarking ML in chemistry is undermined by a hidden confounder: the molecular representation is often treated as a neutral ``input format'', but it actually defines the invariants, and edit operations that models learn over. A typed intermediate representation makes  the syntax explicit, so benchmarking can be done on a stable substrate rather than on brittle serializations. \medskip

Here we explore representation of molecules via Algebraic Data Types (ADTs): composite data types within a programming language, formed from simpler constructors for atoms, connectivity, stereochemistry, and geometry, together with composible operations on that structure. \medskip

After decades of work, string encodings such as SMILES \citep{weininger1988smiles} and, SELFIES \citep{krenn2020self} remain widely used for interchange and as model inputs. They are compact and interoperable, but they primarily serialize a 2D graph-level description and therefore do not directly carry the 3D information (coordinates, conformers) that many tasks ultimately depend on \citep{zheng2019identifying,deng2022artificial}. Coverage of stereochemistry and edge cases varies across toolchains, and complex bonding patterns (general delocalisation, multicentre bonding, many organometallic conventions) are typically handled via additional conventions or specialized tokens rather than a uniform semantic model. For machine learning, a second set of issues matters: token sequences can be syntactically invalid, small token edits can induce large structural changes, and symmetry/invariance structure is not represented as a first-class object. These factors can force models to spend capacity on learning a surface grammar rather than chemistry, complicating controlled benchmarking and interpretation. \medskip

Although SMILES provides a convenient, widely adopted linearization of molecular graphs, it is a highly constrained language: only a small subset of character sequences correspond to syntactically valid, parseable SMILES, and an even smaller subset correspond to chemically meaningful molecules. As a result, any approach that “samples SMILES strings” in the unconstrained sense is dominated by invalid outputs. This is not merely an efficiency nuisance; it fundamentally confounds empirical comparisons, because performance becomes driven by how often a method stumbles into valid syntax rather than by its ability to model molecular structure. For this reason, throughout we treat syntactic validity as part of the modeling/inference problem rather than an after-the-fact filter, and we avoid evaluations that implicitly reward or penalize methods based on arbitrary amounts of invalid-string rejection.

Rather than extending a string grammar until it behaves like a language, as proposed in \citep{krenn2022selfies}, one can treat the molecule itself as a first‑class program object in an ordinary programming language, with explicit types and operations defined on it. In a functional paradigm, ADTs paired with pure functions may provide composability, equational reasoning about transformations on molecules, and immutability, helping to make invariants and exceptions explicit. We propose that if molecules are to be generated, validated, optimised and transformed by programs, representing them directly as typed data -- rather than as tokens that must be decoded -- offers a more transparent and verifiable route. \medskip

In this work, we represent molecules as ADTs so that parsing, validation, and transformation operate on structured molecular objects (atoms, bonds, bonding systems, geometry, and annotations).

Using an ADT turns ``chemical validity'' into explicit invariants attached to constructors and validators. This supports deterministic traversal/feature extraction, localized edits that preserve well-formedness, and principled interoperation layers (import/export) that do not leak toolkit-specific internal conventions into the core representation. \medskip

We first review limitations of common representations (string encodings, fingerprints, and widely used in-memory graph objects) with a focus on expressiveness, edit locality, validation, and compatibility with geometric and Bayesian modeling \citep{bronstein2021geometric,kendall2017uncertainties}. We then introduce our ADT-based representation and its Haskell implementation, including (i) a Dietz-style constitution layer, (ii) a coordinate layer, and (iii) electronic annotations. Finally, we provide examples (e.g., benzene) and a prototype probabilistic-programming integration to illustrate how structured molecular values can be used directly in modeling code. \medskip

A brief note on terminology. We distinguish (i) \emph{storage or transmission} formats that serialise molecules for storage and exchange (e.g., SMILES, InChI, SDF) from (ii) internal data models (\emph{computational data type}) used for editing, featurisation, and learning. In practice, these are sometimes discussed under the same umbrella term “representation”. Making the distinction explicit matters for benchmarking: many properties that ML pipelines rely on—validity conditions, canonicalisation conventions, and the definition of “local edits”—are properties of a data model (plus its validators), not of a character string in isolation. \medskip


Functional programming has been used in cheminformatics primarily as an implementation choice for existing workflows (e.g., leveraging immutability, strong typing, and safer composition) \citep{berenger2019chemoinformatics,hock2012chemf}. Our focus is different: we propose a representation design in which the molecular data model itself is explicitly typed and layered, so that invariants and transformations become part of the representation’s semantics rather than being enforced indirectly by external parsers and toolkit conventions. \medskip

This work presents a reference implementation rather than a final standard. Our goal is to demonstrate how an ADT-based approach can make representational assumptions explicit, support composable transformations, and provide a foundation for future evaluation and extensions in cheminformatics.

Below are some definitions used, defined informally. For formal definitions, see \citep{pierce2002tapl,harper2016pfpl}. \medskip

\begin{definition}[Data type]
A data type is a compile-time contract specifying which values may exist and which operations are meaningful on those values.
\end{definition}
\begin{definition}[Algebraic Data Types (ADTs)]
 An Algebraic Data Type (ADT) is a data type whose possible values are defined by \emph{constructors} (products/records and sums/variants, possibly recursive) and which is deconstructed by \emph{pattern matching}.
\end{definition}

\subsection{Molecular Graphs, hypergraphs, and multigraphs}\label{primitives-theory}

\subsubsection{Molecular Graphs}\label{molecular-graphs}
Many molecular representations are use molecular graphs and thus inherit their strengths and limitations for representation\citep{david2020molecular}. When decoded, the string based representations such as SMILES and SELFIES correspond to molecular graphs, and the atom and bond ``blocks'' of SDF molfiles represent a molecular graph, stored as a connection table.\medskip

\begin{definition}[Graph]
A graph \(G=(V,E)\) is a pair consisting of a set of vertices \(V\) and an edge set \(E \subseteq V \times V\).
\end{definition}

\begin{definition}[Molecular Graph]
A molecular graph is a graph, \(G=(V,E)\), where vertices, \(V\), represent atoms and edges, \(E \subseteq V \times V\), represent bonds. Molecular graphs are connected, undirected, and labeled. Vertex labels may include information such as element or charge, and edge labels may include the bond type as a categorical label, a bond order, or stereochemical annotations \citep{dietz1995yet, david2020molecular}. 
\end{definition}

Graph representations let models exploit chemical graph theory directly. However, because edges relate two vertices only, simple graph encodings can be awkward for delocalized or multicenter bonding patterns and for representations where bond order is not well-defined without additional electronic-structure context \citep{david2020molecular, wigh2022review}. Edge labels (e.g., rational bond orders) do not capture how the \emph{same} electrons contribute to multiple bonds, and encoding delocalisation by multiplying bond types becomes ad hoc. As Dietz noted:
\begin{quote}
``Enhancing the expressiveness by including a new bond type for every exceptional case is certainly not a very elegant solution.''
\end{quote}
A single graph also cannot represent tautomers, typically requiring multiple structures. Hypergraph and multigraph models have been proposed to address some of these issues \citep{kajino2019molecularhypergraphgrammarapplication}, but have seen limited use \citep{david2020molecular}.

\subsubsection{Molecular Hypergraphs}\label{molecular-hypergraphs}
A hypergraph allows edges (hyperedges) to connect any number of vertices. In molecular hypergraphs, 2-vertex hyperedges represent localised bonds; hyperedges with more than two vertices represent delocalised electron systems and can be labeled by the number of shared electrons \citep{dietz1995yet}.

Dietz cautioned that multi-atom hyperedges erase binary neighborhood information:
\begin{quote}
``A hyperedge containing more than two atoms gives us no information about the binary neighborhood relationships between them. That means we have no information at our disposal concerning the way in which the electrons are delocalized over these atoms'' \citep{dietz1995yet}.
\end{quote}
Because multi-vertex hyperedges suppress pairwise adjacencies, they hinder algorithms that rely on binary neighborhood structure, and Dietz therefore rejects hypergraphs for constitutional representation \citep{dietz1995yet}.

\subsubsection{Multigraphs}\label{multigraphs}
A multigraph \(G=(V,E)\) has vertex set \(V\) and a multiset (bag) \(E\) of edges where edges may repeat.

As explained by Dietz\citep{dietz1995yet} Vertices in a molecular multigraph represent atoms, which can be labeled as with molecular graphs, but edges have a different meaning. Each edge is not a bond, but a \emph{bonding system}. 
A single edge between two vertices represents a bond between atoms. Where there are multiple edges between pairs of vertices, each edge represents the atom's share of electrons in bonds in which the atoms participate. 
For instance, in benzene, each C-C pair has two edges, one representing the 2c-2e bond between adjacent carbon atoms, and another edge representing the delocalised elections contributed to the ring. 

By using a multigraph approach, it is possible to label bonds with information about the binary bonding relationships between atoms, as well as describing the delocalised electrons that persist over a subset of atoms \citep{dietz1995yet}.

\subsubsection{Dietz representation}\label{dietz-representation}
Dietz factors structure into constitution (connectivity and electron sharing), configuration (discrete stereochemical arrangements), and conformation (continuous 3D geometry) \citep{dietz1995yet}. In MolADT, we follow Dietz for constitution and supply geometry separately via an explicit coordinate layer; this separation allows a single constitution to be paired with multiple conformers while keeping constitutional invariants stable. \medskip

In MolADT, we adopt Dietz’s constitution layer as the semantic core for bonding (including delocalised and multicentre systems), and we can attach 3D coordinates as a separate layer. This separation supports geometry-free symbolic manipulation when coordinates are absent, while still enabling stereochemical and conformational reasoning when coordinates are present. \medskip

\begin{definition}[Constitution]
The \textbf{constitution} of a molecule is an ordered pair \(C=(V,B)\) where \(V\) is the set of atoms and \(B\) the set of bonding systems:
\[
V=\{\, (u,j,A) \mid u\in\mathbb{N}_0,\ j\in\mathbb{Z}_+,\ A\in\Sigma \,\},
\]
where \((u,j,A)\) represents an atom with \(u\) unshared valence electrons, unique index \(j\), and atomic symbol \(A\in\Sigma=\{H,C,O,\dots\}\).
Let \(J=\{\,j \mid \exists\,u,A:\ (u,j,A)\in V\,\}\) and \(\binom{J}{2}=\{\,\{i,k\}\subset J\mid i\neq k\,\}\). Each bonding system is a pair
\[
B \subseteq \{\, (s,F) \mid s\in\mathbb{N}_0,\ \emptyset\neq F\subseteq \tbinom{J}{2}\,\},
\]
where \(s\) is the number of shared electrons in the system and \(F\) is a nonempty set of unordered pairs of vertex indices (atom pairs) over which those electrons are distributed.
\end{definition}

\paragraph{Examples.}

Intuition before the examples: a ``bonding system'' is an electron pool shared across one or more atom–atom edges. When the pool spans a single edge, it behaves like an ordinary localised bond. When the pool spans several edges, it encodes delocalisation while keeping all pairwise adjacencies explicit. The following examples illustrate both cases: \medskip

Localised bonding: a 2-electron covalent bond between atoms \(i\) and \(j\) is \((2,\{\{i,j\}\})\). Delocalised bonding uses \(|F|>1\), e.g., benzene’s \(\pi\) sextet:
\[
(6,\{\{c_1,c_2\},\{c_2,c_3\},\{c_3,c_4\},\{c_4,c_5\},\{c_5,c_6\},\{c_6,c_1\}\}).
\]

\begin{example}[Constitution of Hydrogen]
\[
V(\mathrm{H}_2)=\{(0,1,H),(0,2,H)\},\qquad B(\mathrm{H}_2)=\{(2,\{\{1,2\}\})\}.
\]
\end{example}

\begin{example}[Constitution of Benzene]
\[
\begin{aligned}
V(\mathrm{benzene})&=\{(0,1,H),\dots,(0,6,H),(0,7,C),\dots,(0,12,C)\},\\
B(\mathrm{benzene})&=\{(2,\{\{1,7\}\}),\dots,(2,\{\{6,12\}\}),\\
&\quad (2,\{\{7,8\}\}),(2,\{\{8,9\}\}),\dots,(2,\{\{12,7\}\}),\\
&\quad (6,\{\{7,8\},\{8,9\},\{9,10\},\{10,11\},\{11,12\},\{12,7\}\})\}.
\end{aligned}
\]
\end{example}

\paragraph{Derived quantities (Dietz).}

The bonding-system representation treats delocalisation as primitive and allows familiar quantities to be derived when needed (e.g., for back-compatibility with bond-order-based tooling). For $b = (s, F)$ we write $n(b) = s$ and $p(b) = |F|$. Dietz defines the electron count associated with an atom $x$ and derived formal charge/bond-order quantities as functions of the bonding systems incident to x. In MolADT these are treated as derived (computed) properties rather than part of the core representation. \medskip

Induced labeled multigraph. For some algorithms it is convenient to work with an induced multigraph view: create an edge $(\{i, j\}, b)$ for each bonding system $b = (s, F)$ and each $\{i, j\} \in F$, labeled by (id(b), s). This makes explicit how a constitution $(V, B)$ yields a multigraph plus metadata grouping edges into shared-electron systems, preserving pairwise adjacency while encoding delocalisation via edge groupings. \medskip

Let \(v(x)\) be the valence of the isolated atom, \(u(x)\) the unshared electrons, \(B_x\) the bonding systems involving atom \(x\), \(n(b)\) the electrons in \(b\), \(p(b)=|F_b|\) the number of atom pairs in \(b\), and \(p_x(b)\) the number of pairs in \(b\) that contain \(x\). Then the electrons belonging to \(x\) in the structure are
\[
v_s(x)\;=\;u(x)\;+\;\sum_{b\in B_x}\frac{n(b)\,p_x(b)}{2\,p(b)}\,,
\]
and the \emph{formal charge} is \(c_s(x)=v(x)-v_s(x)\). The \emph{formal bond order} between \(x\) and \(y\) is
\[
\mathrm{bo}(x,y)\;=\;\tfrac{1}{2}\sum_{b\in B_{\{x,y\}}}\frac{n(b)}{p(b)}\,,
\]
with \(B_{\{x,y\}}\) those bonding systems for which \(\{x,y\}\in F_b\) \citep{dietz1995yet}. (Electrostatic interactions (e.g., ionic or hydrogen-bond contacts) can be modeled, if desired, as bonding systems with s = 0 to record neighborhood relationships without assigning shared-electron counts.) \medskip

\begin{figure}
\centering
\includegraphics[scale=0.15]{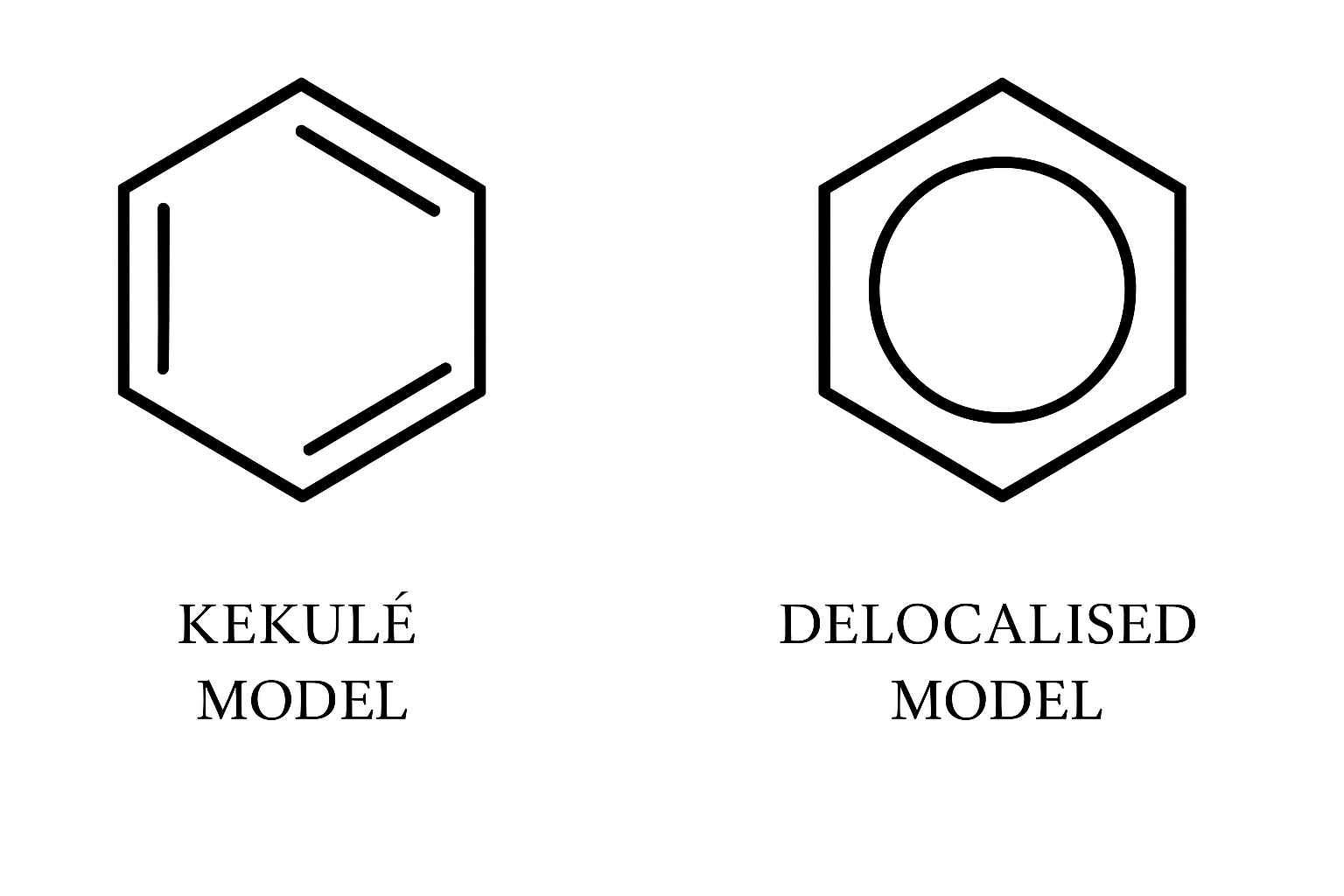}
\caption{Benzene -- Kekul\'e structure on the left and on the right Benzene's more general form. Image Credit: Oliver Goldstein (Author).}
\label{kekule}
\end{figure}

Figure \ref{kekule} contrasts a Kekul\'e depiction with a delocalised depiction of benzene. In Dietz’s formulation, the $\pi$ electrons of the benzene ring are represented as a single bonding system (one electron pool) spanning the six C–C ring edges, separate from the $\sigma$ framework.

The delocalised $\pi$ electrons of the benzene ring are described in a single bonding system, as the final element of the set, and separately from all other covalent bonds in the structure. \medskip

\subsubsection{Beyond benzene: multi-centre and organometallic bonding}

To show that the same constitutional machinery extends beyond typical organic examples, we include an organometallic case. A major advantage of the Dietz representation is its general model for bonding, making it straightforward to describe complex systems such as organometallics (e.g., ferrocene) and electron-deficient bonding (e.g., diborane). It is straightforward to describe complex bonding systems such as organometallics, with ferrocene and diborane already given as motivating examples in the original article. Delocalised bonding is represented explicitly via bonding systems rather than through special atom types or bond labels. This stands in contrast to string-based representations such as SMILES, whose grammar encodes aromaticity through atom typing (e.g.\ \texttt{c}, \texttt{n}, \texttt{[nH]}) and a collection of special-case rules for different heteroaromatic environments. \medskip

One possible Dietz-style constitution for ferrocene can be written schematically as
\[
\begin{aligned}
V(\mathrm{ferrocene})&=\{(0,1,\mathrm{Fe}),(0,2,\mathrm{C}),\dots,(0,11,\mathrm{C}),(0,12,\mathrm{H}),\dots,(0,21,\mathrm{H})\},\\
B(\mathrm{ferrocene})&=\{(2,\{\{12,2\}\}),\dots,(2,\{\{16,6\}\}),(2,\{\{17,7\}\}),\dots,(2,\{\{21,11\}\}),\\
&\quad (2,\{\{2,3\}\}),\dots,(2,\{\{6,2\}\}),(2,\{\{7,8\}\}),\dots,(2,\{\{11,7\}\}),\\
&\quad (6,\{\{1,2\},\dots,\{1,6\},\{2,3\},\dots,\{6,2\}\}),\\
&\quad (6,\{\{1,7\},\dots,\{1,11\},\{7,8\},\dots,\{11,7\}\}),\\
&\quad (6,\{\{1,2\},\dots,\{1,6\},\{1,7\},\dots,\{1,11\}\})\},
\end{aligned}
\]
where the final three bonding systems respectively capture the $\pi$-electron delocalisation within each cyclopentadienyl ring and an illustrative Fe–Cp back-donation pool. The precise choice of bonding systems is not unique, but the representation makes explicit which interactions are localised and which are genuinely multi-centre and delocalised.

\begin{figure}
\centering
\includegraphics[scale=0.4]{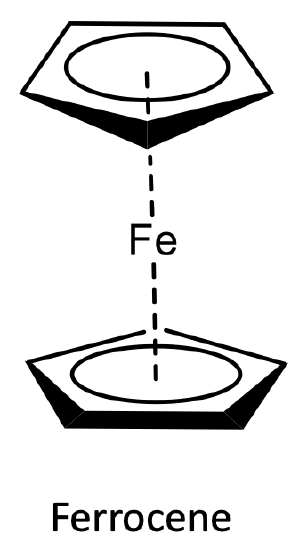}
\caption{Three-dimensional view of \textbf{ferrocene} showing the metallocene ``sandwich'' geometry. Reproduced from \citep{PubChem_Ferrocene_ferrocene_CID129628779}.}
\label{ferrocene-sandwich}
\end{figure}

Classical Lewis theory assumes that bonding can be decomposed into independent two-centre, two-electron (2c–2e) bonds. This assumption already breaks down for electron-deficient and organometallic systems. In ferrocene, bonding is inherently delocalised and multi-centre: the $\pi$ electrons of each cyclopentadienyl (Cp) ring are shared simultaneously among five Fe–C interactions \emph{and} participate in the aromatic C–C bonding within the ring, while the iron centre additionally back-donates $d$ electrons into Cp $\pi^\ast$ orbitals. Any single Lewis-style graph is therefore forced into an unsatisfactory choice: either (i) ten localised Fe–C bonds, which destroys aromaticity and miscounts electrons, or (ii) two abstract “coordinate” Fe$\leftarrow$Cp bonds, which conceal explicit Fe–C adjacency and geometry. \medskip

The Dietz representation avoids this false dichotomy by separating \emph{pairwise adjacency} from \emph{electron delocalisation}. Explicit two-centre edges (Fe–C, C–C, C–H, etc.) are retained to preserve neighbourhood structure and geometry, while delocalised interactions are encoded via labelled bonding systems $(s,F)$—electron pools of size $s$ spanning a finite set of edges $F$. This permits a faithful representation in which Cp aromaticity, Fe–C coordination, and back-donation coexist without inventing artificial bond types or sacrificing chemical semantics. \medskip

Listings~\ref{lst:diborane-adt} and~\ref{lst:ferrocene-adt} provide concrete stress-tests of this approach. Diborane contains two $\mathrm{B\text{–}H\text{–}B}$ bridges that are standardly described as two-electron three-centre (3c–2e) bonds, while ferrocene is characterised by $\eta^5$ (haptic) coordination between the iron centre and each Cp ring.\cite{IUPACGoldBookElectronDeficientBond,Brammer2022TUCAN} In both cases, the localised $\sigma$ framework is expressed as an ordinary adjacency set (\texttt{localBonds}), and each bridge or coordination interaction is represented as an explicit bonding system with a specified electron count distributed over multiple edges. Multi-centre bonding is therefore expressed directly, rather than being squeezed into an ill-fitting 2-centre formalism. \medskip

Despite satisfying the three requirements identified by Krenn et~al.\ for representations beyond organic chemistry—namely support for delocalised bonding, explicit terminal hydrogens, and advanced stereochemical phenomena such as cumulenes—Dietz-style representations are described in that work (apparently mischaracterised as hypergraphs) as leading to “complicated nested sets of brackets that may be hard to comprehend.”\cite{krenn2022selfies} While the set-theoretic notation may be unfamiliar, the underlying constitution of a molecule admits a compact and unambiguous grammar. This stands in contrast to SMILES and SELFIES strings, whose linear grammars rely on overloaded symbols and traversal-dependent conventions to encode bonding semantics.\cite{krenn2022selfies} \medskip 

These limitations are especially acute for organometallic and multi-centre systems. SMILES provides only pairwise bond primitives with a small fixed vocabulary; as a result, such molecules are typically approximated via disconnected charged fragments or dialect-specific conventions (including “zero-order” or haptic workarounds), and encodings need not round-trip or canonicalise consistently across toolchains.\cite{Brammer2022TUCAN,Blanke2025InChIInorganics} SELFIES inherits this restriction because it offers a robust grammar over the same underlying atom–bond graph—guaranteeing syntactic validity but not introducing a native notion of multi-centre electron pools or hapticity.\cite{krenn2020self,Lo2023SELFIESLibrary} 

\subsection{Strings}\label{strings}

\subsubsection{SMILES}\label{smiles}
SMILES encodes a labelled 2D molecular graph as an ASCII string using atoms, bonds, branches, and ring closures, with optional isomeric annotations for stereochemistry. As an interchange format it is compact and widely interoperable, but it does not natively include 3D coordinates or conformational ensembles, and several behaviors depend on toolchain conventions (e.g., aromaticity perception, kekulisation, and canonicalisation) \citep{david2020molecular, wigh2022review,krenn2022selfies,oboyle2012standard,leon2024comparing,mcgibbon2024intuition,kim2021merged}. These factors matter for benchmarking because models trained or evaluated on SMILES inherit representation-specific notions of validity and locality \citep{daylight_manual}. In practice, limitations arise around (i) resonance/tautomerism and delocalisation (which require multiple structures or conventions), (ii) the breadth of stereochemical cases that can be expressed at the graph level, and (iii) variation across toolchains (e.g., aromaticity perception, canonicalisation).\citep{wigh2022review} These are active areas of improvement: OpenSMILES provides a community spec, and newer proposals such as BALSA offer a formally specified subset with explicit syntax to reduce ambiguity across implementations \citep{apodaca2022balsa}. \medskip

While SMILES can represent aromaticity through specific syntax\citep{weininger1988smiles}, it cannot encode delocalized bonding beyond such cases. Resonance and tautomerism must either be simplified into a single structure or require multiple distinct structures to represent each form. \medskip

The encoding of stereochemistry in SMILES has notable constraints. While tetrahedral centers are well-supported through isomeric SMILES, a common extension, the encoding of more complex stereochemical features, such as dynamic or rotational stereoisomerism seen in atropisomers, remains unsupported. Isomeric SMILES can represent certain forms of axial chirality in cumulenes and handle non-tetrahedral stereocenters, including allene-like, trigonal-bipyramidal, and octahedral configurations \citep{daylight_manual}. However, these representations rely on \textbf{2-dimensional} graph-based encoding and lack support for nuanced 3D spatial arrangements required for complex organometallic systems or multi-centre bonding scenarios. The absence of three-dimensional (3D) spatial data further limits SMILES’ utility for applications requiring dihedral angles or precise molecular conformations, a critical aspect of protein-ligand modeling and other machine learning tasks such as property prediction \citep{zhou2022uni,xu2022geodiff,dayalan2006dihedral}. \medskip

Beyond these limitations, SMILES suffers from unique structural and functional shortcomings such as syntactic invalidity \citep{krenn2020self,david2020molecular, wigh2022review}, non-local encoding of features, and non-unique representations of molecules, all of which can impair machine learning models as the models must implicitly learn the SMILES syntax, which can be difficult, and require larger training sets and wasted computation \citep{david2020molecular,fang2022geometry}. \medskip

``Syntactic invalidity'' means that not all valid SMILES strings correspond to valid molecules. In Machine learning contexts, algorithms may output syntactic nonsense, and must re-sample or require larger training sets to ensure robustness \citep{david2020molecular,gomez2018automatic,krenn2020self}. Even though one paper claims that invalid SMILES should be seen as beneficial to chemical language models \citep{skinnider2024invalid}, it should be noted that the perceived improvement is only relative to comparisons between SMILES and SELFIES models, and likely reflects the models' enhanced ability to infer SMILES syntax when generating valid molecules. \medskip

In SMILES, small changes in the string can lead to drastic changes in molecular structure, and vice versa -- adjacent atoms in the molecule can be far apart in the string \citep{fang2022geometry}. As well as impacting readability, the non-local encoding of structural information can make it difficult to write or even find a similarity metric of SMILES strings whose magnitude corresponds to a relevant similarity of molecules \citep{fang2022geometry}. \medskip

SMILES representations are non-unique: a single molecule can be expressed by multiple valid strings \citep{weininger1988smiles}. As well as complicating database searches \citep{wigh2022review}, machine learning algorithms must discern the equality of different representations. Although some machine learning approaches have attempted to utilize multiple SMILES for the same molecule in their training sets and have claimed improvements to their models as a result \citep{li2022novel,arus2019randomized}, these gains may reflect a data-augmentation/regularisation effect: the model is trained to be invariant to multiple serializations of the same underlying graph, but they also underline that the learning objective must spend capacity on the string grammar rather than chemistry. This motivates representations whose locality and invariants are explicit. In other words, it exposes a weakness of SMILES for machine learning, not an advantage. \medskip

\begin{figure}
\centering
\includegraphics[scale=0.25]{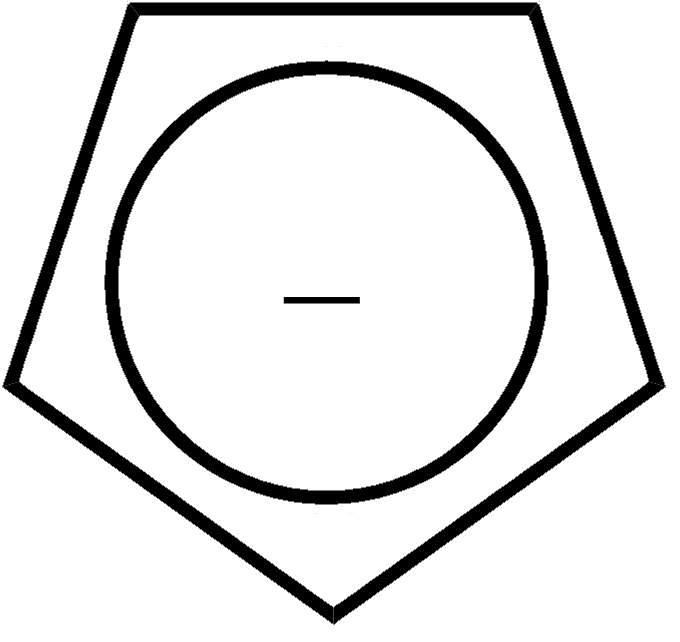}
\caption{The Cyclopentadienyl anion. Although the structure displays clear rotational symmetry, this is not indicated by the SMILES representation: \texttt{[cH-]1cccc1}. Image Credit: Oliver Goldstein (Author).}
\label{cyclo}
\end{figure}

SMILES does not explicitly represent molecular symmetries (graph automorphisms) as first-class objects. For example, the Cyclopentadienyl anion in Figure ~\ref{cyclo} can be written as \texttt{[cH-]1cccc1}, but the rotational symmetry evident in Fig.~\ref{cyclo} is not encoded in the string. Moreover, because SMILES is non-unique, alternative strings such as \texttt{1cccc1[cH-]} represent the same structure without any explicit indication of equivalence at the representation level. \medskip

Attempts to canonicalize SMILES, and the proprietary nature of the Daylight canonicalisation algorithm, have led to inconsistencies between implementations and research teams \citep{david2020molecular}, though the OpenSMILES initiative has mitigated this issue by providing a standardized grammar. \medskip

Importantly, as an untyped data format, a string does not by itself enforce semantic invariants (e.g.\ valence constraints, charge consistency, or well-formed stereochemical annotations); those guarantees live in external parsers, validators, and toolchain conventions. \medskip

While SMILES and its extensions remain useful tools for encoding and sharing molecular structures, its limitations constrain its applicability in modern cheminformatics and machine learning workflows. Extensions to SMILES to fix problems, and the problems they themselves introduce, are discussed in a later section.

\subsubsection{SELFIES}\label{selfies}
SELFIES (Self-Referencing Embedded Strings) \citep{krenn2020self} addresses a major practical issue with SMILES: syntactic invalidity. They create a ``100\% robust'' representation, which would translate to soundness in the terminology of logic or computer science. In particular, they present a Context Free Grammar (CFG) where all sequences of terminal symbols represent molecular graphs (soundness) and all molecular graphs can be represented (completeness -- but only up to the limited CFG they have created, see Fig. \ref{tab:selfies_rules}). The CFG of SELFIES can be translated into sum and product notation as every CFG can be represented by an Algebraic Data Type \citep{syrotiuk1984functional}. \medskip

In attempting to retain semantic as well as syntactic validity after mutation, SELFIES uses a table of derivation rules, with overloading of terms to encode valence information \citep{krenn2022selfies}. Although successful in maintaining semantic validity after mutation (with the given set of rules, for small biomolecules), and as successfully demonstrated with GANs and VAEs\citep{krenn2020self}, the overloading of terms makes the grammar more complicated. The authors claim that SELFIES may be more readable than SMILES for large molecules, although they acknowledge that ``Read-ability is in the eye of the beholder''\citep{krenn2020self}. In practice, SELFIES tokens can be less directly interpretable than SMILES substrings, because the mapping from tokens to familiar chemical motifs is less transparent; whether this matters depends on the use case (human inspection vs model robustness). \medskip

For machine learning, SELFIES still presents a token sequence governed by a formal grammar. Models therefore must learn token-level regularities to generate valid structures, even if the decoder guarantees validity for any terminal sequence \citep{david2020molecular}. Claims of improved ``machine readability'' should be interpreted cautiously: whether robustness yields systematic gains depends on the task, architecture, and evaluation protocol. \medskip

In light of earlier discussions of line-notation design trade-offs \citep{weininger1988smiles}, the robustness -- interpretability balance in SELFIES is best viewed as a pragmatic engineering choice: additional rule structure can improve validity under mutation but may reduce human interpretability compared with simpler notations. \medskip

\begin{table}[h]
    \centering
    \tiny
    \renewcommand{\arraystretch}{1.2}
    \setlength{\tabcolsep}{2pt}
    \begin{tabular}{c|c|c|c|c|c|c|c|c|c|c|c|c|c|c}
        \toprule
        State & $\epsilon$ & [F] & [=O] & [\#N] & [O] & [N] & [=N] & [C] & [=C] & [\#C] & [Branch1] & [Branch2] & [Branch3] & [Ring] \\
        \midrule
        $X_0$ & $X_0$ & F & 0 & $X_2$ & N & $X_3$ & 0 & $X_2$ & $N X_3$ & $C X_4$ & $C X_4$ & $C X_4$ & ign $X_0$ & ign $X_0$ \\
        $X_1$ & $\epsilon$ & F & 0 & N & 0 & $X_1$ & $N X_2$ & $N X_2$ & $C X_3$ & $C X_3$ & ign $X_1$ & ign $X_1$ & R(N) & R(N) \\
        $X_2$ & $\epsilon$ & F & =O & $N X_2$ & 0 & $X_1$ & $N X_2$ & =N $X_1$ & =C $X_2$ & =C $X_2$ & B(N, $X_5$) $X_1$ & B(N, $X_5$) $X_1$ & R(N) $X_1$ & R(N) $X_1$ \\
        $X_3$ & $\epsilon$ & F & =O & \#N & 0 & $X_1$ & $N X_2$ & =N $X_1$ & =C $X_2$ & \#C $X_1$ & B(N, $X_5$) $X_2$ & B(N, $X_6$) $X_1$ & R(N) $X_2$ & R(N)$X_2$ \\
        $X_4$ & $\epsilon$ & F & =O & \#N & 0 & $X_1$ & $N X_2$ & =N $X_1$ & =C $X_2$ & \#C $X_1$ & B(N, $X_5$) $X_3$ & B(N, $X_7$) $X_1$ & R(N) $X_3$ & R(N)$X_3$ \\
        $X_5$ & C & F & 0 & N & 0 & $X_1$ & $N X_2$ & $N X_2$ & $C X_3$ & $C X_3$ & $X_5$ & $X_5$ & $X_5$ & $X_5$ \\
        $X_6$ & C & F & =O & $N X_2$ & 0 & $X_1$ & =N $X_1$ & =C $X_2$ & =C $X_2$ & $X_6$ & $X_6$ & $X_6$ & $X_6$  & $X_6$ \\
        $X_7$ & C & F & =O & \#N & 0 & $X_1$ & =N $X_1$ & =C $X_2$ & \#C $X_1$ & $X_7$ & $X_7$ & $X_7$ & $X_7$  & $X_7$  \\
        $N$ & 1 & 2 & 3 & 4 & 5 & 6 & 7 & 8 & 9 & 10 & 11 & 12 & 13 & 14 \\
        \bottomrule
    \end{tabular}
    \caption{Derivation rules of SELFIES \citep{krenn2020self} for molecules in the QM9 dataset.}
    \label{tab:selfies_rules}
\end{table}

Again, like SMILES, resonant structures are not directly expressible in SELFIES, and delocalized bonding (other than aromaticity) cannot be expressed, along with tautomers. \medskip

Since publication, SELFIES has added the ability to represent chirality, through importation of symbols from SMILES, although some forms produce errors in decoding. However, three dimensional information is not included, and it is unclear how to extend the grammar whilst retaining its simplicity and soundness. The spatial arrangement of ligands in square-planar stereochemistry, like that of cisplatin and transplatin, cannot be fully captured without 3D information, and so in exploring the chemical space, SELFIES remains unable to represent potentially crucial information for novel drug discovery and property prediction, despite being sound. \bigskip

Furthermore, E-Z stereochemistry does not in general provide accurate dihedral or torsional angles. Proteins rely on these angles for many tasks, which may be critical for machine learning tasks. As Dayalan explains (\citep{dayalan2006dihedral}) ``Dihedral angles are of considerable importance in protein structure prediction as they define the backbone of a protein, which together with side chains define the entire protein conformation.''. Angles influence the energy associated with a conformation of a drug which in turn relates to the utility and stability of the drug \citep{anderson2003process}. Although SELFIES solves the syntactic invalidity problem of SMILES, \textit{many} challenges remain. 

\subsubsection{Extensions}\label{extensions-brief}
\paragraph{The future of SELFIES and Future Projects}
In ``SELFIES and the Future of Molecular String Representations'' Krenn et al. (2022) \citep{krenn2022selfies}, review the limitations of string-based molecular representations in cheminformatics and propose future directions, primarily centered around extending string-based methods. \medskip

They highlight several representational difficulties with current molecular string formats, including their inability to effectively capture macromolecular structures, crystal lattices, complex bonding (e.g., organometallics), advanced stereochemistry, precise conformations, and non-covalent interactions such as hydrogen and ionic bonding. These limitations, they argue, are not only present in SMILES and SELFIES but are features of most modern, digital molecular representations in use. \medskip

The authors also identify challenges specific to machine learning applications, such as issues arising from the non-local encoding of molecular features within strings, and the impact this has on the latent space of VAEs, as well as the impact overloading of symbols has the efficiency of generative models, as well as human readability. 

\paragraph{Macromolecules and crystals.} \medskip
For macromolecules, Krenn et al. discuss existing approaches such as CurlySMILES, BigSMILES, and HELM, each of which introduces distinct syntactic modifications to account (or not, as the case may be) for repeating subunits, complex connectivity, and stochastic relationships between subunits. They propose BigSELFIES and HELM-SELFIES as future extensions to enable the use of generative models which take strings as input for macromolecular synthesis and design. \medskip
 
In the case of crystals, the authors propose yet another syntax, Crystal-SELFIES, based on the labeled quotient graph of the crystal structure. The formalism of the approach is largely unexplored in the paper, however relies on representation of subunits, in a related manner as with macromolecules. 

\paragraph{Extensions of SMILES.}
SMILES remains the dominant line notation, but fixes and extensions continue to appear. DeepSMILES reduces bracket/closure errors yet does not eliminate syntactic invalidity altogether \citep{o2018deepsmiles}. Daylight’s ecosystem introduced related languages (e.g., SMARTS/SMIRKS for queries and transforms; CHUCKLES/CHORTLES for mixtures), and OpenSMILES provides a community specification \citep{daylight_site}. For polymers, BigSMILES extends string notation to stochastic connectivity \citep{david2020molecular}. For reactions, SMIRKS captures structural transforms but leaves conditions to external metadata; later, RInChI added a layered, direction‑aware identifier, and ProcAuxInfo was proposed for process data (yields, temperature, concentrations) \citep{doi:10.1021/acs.jctc.8b00640,wigh2022review,heller2013inchi}. These ad-hoc improvements continue to this day \citep{Reboul2025}. Finally, Balsa offers a fully specified, machine‑readable subset of SMILES to reduce ambiguity across implementations \citep{apodaca2022balsa}. Extensions to SMILES/SELFIES continue to improve coverage (polymers, reactions, subsets with stricter semantics, etc.), but they also fragment the representational landscape: each extension introduces additional syntax rules and often new categorical labels, making round-trips and benchmark comparisons harder across toolchains and datasets. For ML evaluation, this matters because “validity”, augmentation strategies, and edit locality become representation- and implementation-dependent. Our focus is therefore not to propose yet another line notation, but to define a typed semantic core (MolADT) with explicit invariants and transformations, and to treat string formats as serialisations at the boundary.

\paragraph{Molecular fingerprints}

Fingerprints are widely used as fixed-length descriptors in classical QSAR and virtual screening benchmarks, and they highlight an important distinction for evaluation: a fingerprint is not a primary molecular representation, but a derived feature map from an underlying representation (graph/string/toolkit object). We therefore discuss fingerprints briefly to clarify where MolADT fits: MolADT is intended as the semantic substrate from which descriptors (including fingerprints) can be deterministically derived, rather than as a competing descriptor family. \medskip

Fingerprints encode molecules into fixed-length feature vectors for retrieval, clustering, and QSAR. They are effective as \emph{descriptors}, but can be lossy with respect to stereochemistry and geometry and are not, by themselves, canonical molecular \emph{representations}. In this paper we therefore treat fingerprints as downstream descriptors derived from a structured representation, not as the representation itself. These approaches may be effective for retrieval and similarity ranking, but they can be lossy with respect to stereochemistry or geometry, and they often require careful choices of similarity metrics and calibration when used for learning or uncertainty estimation \citep{capecchi2020one, boldini2024effectiveness, bajusz2015tanimoto}. In this paper we treat fingerprints as downstream descriptors rather than as a primary molecular representation. Classical examples include atom pairs and topological torsions (topology), ECFP/FCFP (local neighborhoods), and pharmacophore pairs/triplets (functional interactions); string‑based variants operate directly on SMILES (e.g., LINGO; MAP4) \citep{boldini2024effectiveness,capecchi2020one}. In practice, analysis relies on set/bit similarities rather than \textbf{vector‑space algebra}; the Tanimoto (Jaccard) index is standard \citep{bajusz2015tanimoto}. Limitations are well known: tautomers may hash to different bitmaps \citep{martin2009let}, and hashing plus high dimensionality can induce bit collisions and sparsity that limit expressivity \citep{fang2022geometry}.  \medskip

\subsection{Existing Programming language representations:}\label{existing-pl-reps}

\subsubsection{Molecules and programming languages}\label{molecules-pl}
Inductive Logic Programming (ILP) has represented molecules, notably in Srinivasan et~al. \citep{srinivasan1996theories}, but the emphasis was rule induction on graphs rather than ILP as a general molecular representation, which likely limited uptake and saw it omitted from recent surveys such as Wigh et~al. \citep{wigh2022review}. In this line, molecules are encoded as Prolog rules in a first-order declarative language; yet Prolog's narrow primitive types and operational features, including negation as failure and side-effecting built-ins like \texttt{read}, \texttt{get}, and \texttt{assert}, mean it is not purely declarative and can complicate reasoning. Srinivasan et~al. used Progol, an ILP extension, to induce rules \citep{srinivasan1996theories} (distinct from ProbLog, a probabilistic logic programming language). More broadly, de~Meent et~al. argue modern AI progress stems from tools (e.g., NumPy) that automate gradients, extended by probabilistic programming to generic modeling with uncertainty \citep{van2018introduction}; early ILP work did not integrate such machine-learning techniques \citep{srinivasan1996theories}.

\subsubsection{Functional languages and cheminformatics}\label{functional-cheminf}
Work using functional programming in cheminformatics has been done before, however the focus of such efforts were usually related to ease of writing or safety of code, or efficiency of functional programming languages, rather than attempting to introduce a new representation by explicitly considering data types.\medskip

Ouch \citep{OuchGitHub} is a chemical informatics toolkit written entirely in Haskell, released in 2010. It was released ``in the hope that it will be useful'', but has seen little development since 2013. Internally, molecules are represented with a molecular graph, implemented as an Algebraic Data Type, where bond types are categorically labeled (e.g. ``sigma'', ``pi'', or ``aromatic''), and without any support for stereochemical features or 3D information.\medskip

Chem\textsuperscript{f}\citep{hock2012chemf} is a toolkit written in Scala. Its purpose was not to create a molecular representation, but to permit existing cheminformatics workflows to utilise the functional programming paradigm, and to avoid the problems which can occur with mutable state and null pointers. The representation used is a 2D molecular graph written as a connectivity list.  \medskip

``Radium'' \citep{Langner2017Radium} is a simple Haskell library. It has support for atomic orbital information, but molecules may only be encoded as a linear notation based upon SMILES strings. Interestingly, the data types used for molecular representation are algebraic, and while being based upon SMILES syntax, are not in fact ``strings''.\medskip

Other work \citep{berenger2019chemoinformatics} has suggested using OCaml for cheminformatics, however this work was primarily written as an introduction to functional programming for chemists, with a granular explanation of specific functions, and a focus on ease of writing code, type safety, and efficiency, such as parallelization. \medskip

To our knowledge, most existing cheminformatics representations—regardless of implementation language—treat the molecular graph (with categorical bond labels) as the semantic core, with validation and special cases handled in library code. In contrast, MolADT is explicitly designed as a typed, layered semantic IR with first-class support for general bonding systems and well-scoped transformations; related functional-language efforts typically reimplement conventional graph models rather than rethinking the representation’s semantic contract.

\subsubsection{RDKit as a Representation}\label{rdkit-repr}
RDKit is a mature and widely adopted cheminformatics toolkit that provides robust parsing, editing, and interoperability. We do not aim to replace RDKit’s algorithms. Our focus is representation-level: we propose a semantic core intended for controlled benchmarking and for workflows where explicit invariants, compositional transformations, and symmetry-aware modelling are first-class concerns. \medskip

Conceptually, RDKit’s internal model is a mutable labelled graph with 3D geometry stored separately (e.g., in conformer objects \citep{rdkit-api-docs}), and many chemical interpretations (aromaticity, delocalisation, stereochemistry) are encoded via enumerated labels and toolkit-specific conventions [46]. This design is practical and widely successful, but it can make it harder to treat molecules as immutable mathematical objects with a single explicit set of invariants spanning constitution and geometry. MolADT instead encodes constitution, optional geometry, and optional annotations in a single typed value with explicit validation and local edits. \medskip

In this paper we use RDKit only as a point of comparison (and, where needed in future work, as an interoperability boundary); MolADT is the semantic representation under study.\medskip

\section{Methodology}

\subsection{Our Algebraic Data Type}

\textbf{A typed molecular representation.} \newline

We implement MolADT in Haskell, a statically typed, lazy functional programming language with native support for algebraic data types and pattern matching \citep{hudak1990haskell, damas1982principal,pierce2002tapl,wright1994syntactic}. This choice is pragmatic: it allows us to (i) express molecules as structured values with an explicit, checkable shape, (ii) implement validation and transformations as composable functions, and (iii) make the presence or absence of optional layers explicit to downstream code. The representation itself is language-agnostic; Haskell provides a concise reference implementation that makes the design concrete. Two language features are central to our design.  First, \emph{type classes} provide named bundles of operations that a type implementing such typeclass promises to support, i.e. \emph{instances} supply the implementations; this yields principled operator overloading and generic code \citep{wadler1989typeclasses}. Second, \emph{static typing} checks these contracts before execution.

\textbf{Constitution (Dietz valence‑multigraph).}\newline
At the core, we adopt the Dietz valence‑multigraph for molecular constitution \citep{dietz1995yet}. By construction, it supports delocalised and multicentre bonding, organometallic and electron‑deficient systems, resonance alternatives, fractional/zero bond orders, ionic bonding, and explicit hydrogens. This generality addresses three limitations identified for extending SELFIES beyond organic chemistry delocalisation, explicit hydrogen counts, and complex stereochemistry \citep{krenn2022selfies}. The constitution layer exposes class‑constrained constructors and pattern matches: instances capture admissible compositions (e.g., what counts as a bond or atom in a given sublanguage), while static typing rules out structurally impossible assemblies before runtime.

\textbf{Coordinates}\newline
A coordinate layer attaches three‑dimensional atomic positions to the constitutional graph. When present, stereoisomers are distinguished and geometric features (bond lengths, angles, torsions) are computable with modest additional storage. The layer is strictly separated from constitution to enable geometry‑free symbolic manipulation. Type‑class‑based interfaces present the same traversal and query operations to both geometry‑free and geometry‑aware values, while static types make the presence/absence of coordinates explicit to clients of the API.

\textbf{Electronic structure (\texttt{Orbital} ADT)}\newline
We implement a dedicated \texttt{Orbital} ADT to model shells, subshells, and orbitals with occupancy. Conceptually, it factors into (i) principal shell, (ii) subshell type (\texttt{s}, \texttt{p}, \texttt{d}, \texttt{f}, \dots), (iii) orbital index within a subshell, and (iv) electron population (integer or fractional). The \texttt{Orbital} ADT can be associated with atoms or bonds, enabling explicit representation of electron counts, lone pairs, and multicentre electron sharing. Because \texttt{Orbital} is first‑class, the same fold/unfold discipline applies. Here, type classes provide the admissible operations (e.g., population queries, spin projections), while instance‑guided \emph{smart constructors} and static types constrain construction to chemically valid states (e.g., occupancy ranges), catching errors early.

\textbf{Type driven composition and traversal.}\newline
Haskell type classes specify admissible molecular components and compositions, constraining construction of ill‑typed structures. Standard higher‑order and monadic combinators (e.g., \texttt{map}, \texttt{fold}, \texttt{forM\_}) then provide idiomatic iteration, filtering, and stateful exploration over molecules, coordinates, and orbitals. This combination type class abstraction plus static typing and inference yields reusable, strongly typed building blocks for downstream algorithms without sacrificing genericity.

\textbf{Reactions and probabilistic programs.}\newline
The same typed representation extends to elementary reactions by lifting the constitution/coordinate/orbital layers over multisets of molecules with atom‑mapping annotations. Type classes define the reaction‑level interfaces (e.g., how to apply an atom map, how to aggregate stoichiometry), and static typing ensures that these lifts preserve invariants across reactants and products. To illustrate compatibility with Bayesian machine learning, we provide a reference implementation of a probabilistic model in Lazy PPL that consumes/produces values of the ADT. These implementations are architectural: they define the data, interfaces and implementations used by experiments in later sections.

\textbf{Practical role.}\newline
Beyond in‑memory computation, the ADT serves as a stable, serialisable format for storage, retrieval, and exchange of molecular information. Because molecules are constructed directly as well‑typed programs in the ADT’s grammar, learning and inference can operate on the representation itself without ad hoc encodings or external context. Type classes deliver uniform tooling (pretty‑printing, hashing, equality), while static typing and inference deliver early error detection and refactor‑friendly guarantees that are essential for reliable cheminformatics pipelines.

\subsection{Haskell}

In Haskell, \texttt{data} defines Algebraic Data Types (ADTs) as closed sets of constructors, so every value has a known shape and functions can pattern‑match exhaustively. Type classes declare which operations a type supports, together with intended laws that instances should satisfy enabling ad‑hoc polymorphism (e.g., \texttt{Eq} for \texttt{==}). Optionality is explicit via \texttt{Maybe}/\texttt{Nothing} (e.g., a missing \texttt{SubShell}), avoiding nulls and exception driven control flow. Haskell’s static types and purity shift many failures to compile time, which is valuable for scientific ML pipelines. \medskip

Our contribution is to bring these guarantees to probabilistic programming for cheminformatics. Existing Haskell Probabilistic Programming Languages (PPLs) (e.g., LazyPPL) support concise Bayesian models where lazy evaluation defers costly sampling/simulation until needed useful for molecular property modeling and design \citep{scibior2015practical,nguyen2022modular,paquet2021lazyppl,mervin2021uncertainty,moss2020gaussian,semenova2020bayesian}. The ADT layer integrates naturally with robust graph/string encodings used in de novo design, such as SELFIES and DeepSMILES \citep{krenn2022selfies,o2018deepsmiles}. (ADTs and purity are not unique to Haskell; Haskell simply offers a mature, succinct realization.)

\subsection{Record Syntax}

In our reference implementation, atoms are typed records (\texttt{Atom\{atomID, atomicAttr, coordinate, shells\}}). Named fields give local, direct access to exactly the features needed for ML or simulation (e.g., \texttt{symbol (atomicAttr a)}, \texttt{x (coordinate a)}), avoiding whole‑molecule string parsing and touching only the required data -- lighter than non‑local encodings such as SMILES/SELFIES \citep{krenn2022selfies,o2018deepsmiles}. The model can evolve by adding fields (e.g., charges, isotopes) with minimal impact when code uses named fields. \bigskip

\begin{minipage}{\linewidth}
\begin{lstlisting}[label={lst:set},language=Haskell,caption={Updating coordinates: structure-preserving local edit}]
updateXCoordinate :: Double -> Atom -> Atom
  updateXCoordinate newX a =
  a { coordinate = fmap (\(Coordinate _ y z) -> Coordinate newX y z) (coordinate a) }
\end{lstlisting}
\end{minipage}

\emph{Explanation:} The pattern destructures the nested \texttt{Coordinate}, replaces \texttt{x} with \texttt{newX}, and reconstructs the outer \texttt{Atom}; other fields are passed through unchanged. This could also return a part of the structure. \medskip

\subsection{Use of AI-assisted editing}
We used OpenAI’s GPT-5.2 Pro (ChatGPT) only for light editorial assistance (copy-editing, LaTeX hygiene, float placement). All domain content mathematics, chemistry, Haskell code, experiments, and conclusions was written by the authors; some code refactoring drew on OpenAI’s Codex. Any LLM suggestions were edited, verified, and cited. No proprietary or human-subject data were provided. The authors take full responsibility for the manuscript and codebase.

\section{Reference Implementation and Worked Examples}
\subsection{The Molecule Algebraic Data Type}
\begin{lstlisting}[
  language=Haskell, style=HaskellStyle,
  caption={Core molecular ADT: Dietz-style constitution with explicit atoms, $\sigma$-adjacency, and electron-pool bonding systems.},
  label={lst:mol-adt-core},
  floatplacement=htbp
]
newtype AtomId     = AtomId Integer deriving (Eq, Ord, Show, Read)
newtype SystemId   = SystemId Int    deriving (Eq, Ord, Show, Read)
newtype NonNegative = NonNegative { getNN :: Int } deriving (Eq, Ord, Show, Read)

-- Canonical undirected edge (store once; enforce i<=j in ctor)
data Edge = Edge AtomId AtomId deriving (Eq, Ord, Show, Read)

data BondingSystem = BondingSystem
  { sharedElectrons :: NonNegative, memberEdges :: Set Edge } deriving (Eq, Show, Read)

data ElementAttributes = ElementAttributes
  { symbol :: AtomicSymbol, atomicNumber :: Int, atomicWeight :: Double } deriving (Eq, Show, Read)

data Atom = Atom
  { atomID :: AtomId, attributes :: ElementAttributes, coordinate :: Coordinate, formalCharge :: Int }
  deriving (Eq, Show, Read)

data Molecule = Molecule
  { atoms :: Map AtomId Atom, localBonds :: Set Edge, systems :: [(SystemId, BondingSystem)] }
  deriving (Eq, Show, Read)
\end{lstlisting}

In Listing \ref{lst:mol-adt-core}, the triple $\langle$atoms, $\sigma$‑adjacency, bonding systems$\rangle$ mirrors Dietz’s constitution $C=(V,B)$: pairwise neighbourhood is preserved by explicit undirected edges, while delocalised or multicentre bonding is captured by pooling $s$ shared electrons across a \emph{set of edges}; this avoids ad hoc bond ``types'' yet keeps binary adjacencies available to algorithms. \medskip

Chemically meaningful quantities are derived rather than hard‑coded: formal bond orders and per‑atom electron counts are obtained by summing each system’s fractional share over its member edges, so the record stays minimal but interpretable. \medskip

Canonical undirected edges and ordered maps/sets make the structure deterministic and easy to validate (idempotent insertions, unambiguous lookups). \medskip

These design choices follow Dietz’s rationale and examples (benzene, diborane, ferrocene), which advocate ``bonding systems'' spanning multiple pairs while retaining pairwise information. \medskip

An \texttt{Atom} records its stable identifier (\texttt{atomID :: AtomId}), immutable elemental metadata (\texttt{attributes :: ElementAttributes}), a Cartesian \texttt{coordinate :: Coordinate}, electronic \texttt{shells :: Shells}, and an explicit \texttt{formalCharge :: Int}. Elemental metadata comprises \texttt{symbol :: AtomicSymbol}, \texttt{atomicNumber :: Int}, and \texttt{atomicWeight :: Double}. Coordinates are expressed in \AA\ via the units type \texttt{Angstrom}, ensuring unit safety for downstream geometry. The \texttt{Shells} type is re‑exported from the orbital layer to allow atoms to carry ground‑state (or user‑specified) configurations. Because \texttt{localBonds} is a set, adding the same edge twice is idempotent and cannot inflate degree artificially.

\begin{lstlisting}[label={lst:dietz-primitives},language=Haskell,
  caption={Dietz constitution primitives used by the ADT. \texttt{mkEdge} canonicalises undirected pairs; \texttt{BondingSystem} pools \(s\ge 0\) shared electrons over a set of member edges.}]
newtype AtomId   = AtomId Integer deriving (Eq, Ord, Show, Read)
newtype SystemId = SystemId Int    deriving (Eq, Ord, Show, Read)

-- Non-negative electron counts for Dietz pools
newtype NonNegative = NonNegative{getNN :: Int} deriving (Eq, Ord, Show, Read)

-- Canonical undirected edge: the constructor enforces ordering
data Edge = Edge AtomId AtomId deriving (Eq, Ord, Show, Read)

-- One Dietz bonding system: s shared e- over a set of edges, with an optional tag
data BondingSystem = BondingSystem
  { sharedElectrons :: NonNegative
  , memberAtoms     :: Set AtomId
  , memberEdges     :: Set Edge
  , tag             :: Maybe String
  } deriving (Eq, Show, Read)

\end{lstlisting}

Listing \ref{lst:dietz-primitives} realises the Dietz constitution \(C=(V,B)\) in a typed record: \texttt{atoms} is the vertex set \(V\); \texttt{localBonds} holds canonical undirected \(\sigma\)-adjacencies; and \texttt{systems} is a finite family \(B=\{(s,E)\}\) of \emph{bonding systems}, each pooling \(s\!\ge\!0\) shared electrons over a nonempty set \(E\) of atom–atom edges. Pairwise neighbourhoods remain explicit (via edges) while delocalised or multicentre effects are handled uniformly by the electron pools -- precisely the computer‑oriented counterpart of structural formulas advocated by Dietz.

\medskip
\noindent
\textbf{Smart constructors and invariants.}
Updates pass through small ``smart'' constructors:
\emph{(i)} \texttt{mkEdge} canonicalises undirected pairs, so set-theoretic semantics are by construction (idempotent insertion, unambiguous lookup);
\emph{(ii)} \texttt{mkBondingSystem} takes a pool size \(s\in\mathbb{N}_0\) and an edge set \(E\), derives the cached atom scope \(\mathrm{atoms}(E)\), and rejects ill‑formed inputs (negative \(s\), empty \(E\), or non‑canonical edges).
Despite, potentially being overridden, maintaining the use of these constructors ensures the invariant \(\mathrm{memberAtoms}=\mathrm{atoms}(\mathrm{memberEdges})\) and keeps \texttt{systems} a well‑formed finite family, in line with the paper’s requirement that bonding systems be defined over sets of atom pairs.

\medskip
\noindent
The record stores primitives; chemically useful quantities are \emph{derived}:
per‑atom electron use \(e(v)\) and per‑edge effective order \(\mathrm{order}(e)\) come from summing each system’s fractional share over its member edges, rather than hard‑coding fractional labels. Coordinates attach configuration/conformation without changing constitution. Deterministic containers (\texttt{Map}/\texttt{Set}) plus canonical edges make validation and permutation‑invariant algorithms straightforward, and the Dietz pooling formalism scales from localised bonds to general delocalisation without inventing ad‑hoc bond types. 

\subsection{Example: Benzene}

The same molecule can be specified in a considerably more compact functional style. However, that approach introduces advanced idioms that trade clarity for brevity. Because the aim here is pedagogical transparency as well as correctness, we retain the longer, fully explicit representation in Appendix~\ref{lst:benzene-adt}.

\noindent\textbf{How to read Appendix~\ref{lst:benzene-adt} (Dietz constitution, step‑by‑step).}
\begin{enumerate}
  \item \emph{Name the atoms \& templates.} The code first fixes stable identifiers \(\texttt{AtomId}\ 1\ldots12\) and caches metadata (\texttt{elementAttributes}, \texttt{elementShells}) for C \& H.
  \item \emph{Build atoms with 3D coordinates.} Records \(\texttt{c1}\ldots\texttt{c6}\) and \(\texttt{h7}\ldots\texttt{h12}\) are constructed with IDs, element attributes, Ångström coordinates, shells, and \(\texttt{formalCharge}=0\).
  \item \emph{Link IDs \(\to\) records.} \texttt{atomTable} is a \(\texttt{Map}\ \texttt{AtomId}\!\to\!\texttt{Atom}\) assembled by successive \(\texttt{M.insert}\), providing fast lookup of each atom by its symbol/ID.
  \item \emph{Lay down the \(\sigma\) framework.} \texttt{sigmaFramework} is a \(\texttt{Set}\) of undirected edges (\(\texttt{mkEdge}\)) for the six C–C ring connections and six C–H bonds this is the pairwise adjacency.
  \item \emph{Add the Dietz \(\pi\) system.} \texttt{piRingEdges} repeats the ring’s C–C edges; \texttt{piRingSystem = mkBondingSystem (NonNegative 6) piRingEdges} creates one Dietz electron pool with \(s=6\) shared electrons delocalised over those edges.
  \item \emph{Assemble the molecule.} \texttt{Molecule \{ atoms = atomTable,\ localBonds = sigmaFramework,\ systems = [(SystemId 1, piRingSystem)] \}}.
\end{enumerate}

\noindent\textit{Interpretation (Dietz).} The triple \(\langle\)\texttt{atoms}, \(\sigma\)-\texttt{localBonds}, \texttt{systems}\(\rangle\) mirrors Dietz’s constitution \(C=(V,B)\): explicit edges preserve pairwise neighbourhoods, while a single \(\pi\)-pool (\(s=6\)) captures delocalisation across the ring without inventing special bond types. Coordinates are attached for geometry but do not change the constitution.

\subsubsection{Validator}\label{subsec:validator}

We validate every molecule immediately after construction or import, so downstream inference and learning code never explores structurally nonsensical states. The validator enforces three structural invariants (i) every bond endpoint must refer to an existing atom, (ii) self-bonds are disallowed, and (iii) the internal bookkeeping for undirected connectivity is symmetric and then applies a conservative, element-wise valence bound. Concretely, each atom’s ``valence usage'' is computed from the local $\sigma$ adjacency together with the fractional contributions implied by any Dietz bonding pools, and the molecule is rejected if any atom exceeds the element-specific maximum (\texttt{getMaxBondsSymbol}). Successful validation is a no-op (the molecule is returned unchanged); failures return a short diagnostic describing which invariant was violated. The routine runs in time linear in the number of atoms and edges.
\subsection{Orbital ADT}

\textbf{Physics-aware data and validation (Aufbau, Hund, Pauli).} \newline
In our reference implementation, quantum information is exposed through simple ADTs rather than dependent or refinement types. Orbitals, subshells, and shells are concrete datatypes (\texttt{Orbital}, \texttt{SubShell}, \texttt{Shell}) with integer occupancies and orientation/hybrid terms; atoms carry a \texttt{shells :: Shells} field inside the molecular record (modules \texttt{Orbital} and \texttt{Molecule}; see \texttt{src/Orbital.hs} and \texttt{src/Molecule.hs}). Client code does not insert electrons via an API that enforces Aufbau/Hund/Pauli at the boundary; instead, it typically obtains ground-state occupancies via a single mapping \texttt{elementShells :: AtomicSymbol -> Shells} and associated element tables (\texttt{elementAttributes}, \texttt{getMaxBondsSymbol}) in \texttt{Constants} (see \texttt{src/Constants.hs}).

Accordingly, the interface provides faithful data structures and convenient constructors/defaults; the implementation supplies the empirical per-element configurations and bond limits as tables. Runtime physics checks presently cover only generic structural validity (e.g., symmetric bonds and per-element maximum total bond order) via \texttt{validateMolecule} in \texttt{Validator} (see \texttt{src/Validator.hs}); they do not currently re-enforce Aufbau/Hund/Pauli for arbitrary edits to \texttt{electronCount}. We therefore document Aufbau/Hund/Pauli as invariants and provide ground-state defaults (with room to extend to ions/exceptions), while leaving strict enforcement to construction discipline or future smart constructors~\citep{iupac_goldbook_aufbau,IUPAC2025Hund,IUPAC2025Pauli}.

\textbf{Hybridisation in the ADT.}\newline
Hybrid orbitals are represented explicitly via
\texttt{hybridComponents :: Maybe [(Double, PureOrbital)]}, a linear combination of pure $s/p/d/f$ orbitals. We treat the coefficients as mixing amplitudes whose squares sum to 1 (checked by smart constructors), and use \texttt{orientation} to store the spatial axis of the resulting hybrid. This captures common hybrids (e.g., $sp^3$, $sp^2$) and supports chemically meaningful predictions such as how $s$‑character modulates acidity and bond geometry (Bent's rule and modern NBO analyses)~\citep{alabugin2015hybrid,weinhold2005nbo}. \medskip

\subsection{Reactions}

\textbf{Reaction ADT} \newline
The proposed ADT can be extended to model chemical reactions. By defining a \texttt{Reaction} data type that captures the reactants, products, and conditions under which a reaction occurs, we can create a tool for studying and simulating chemical processes.\medskip

Listing \ref{lst:reaction-representation} illustrates a \texttt{Reaction} data type, where reactants and products are represented as lists of pairs, containing a stoichiometric coefficient (a \texttt{Double}) and a \texttt{Molecule}. The \texttt{conditions} field captures reaction conditions such as temperature and pressure, defined with respect to a time range. Listing \ref{lst:reaction-example} provides an example reaction, illustrating the formation of water from hydrogen and oxygen under specific thermodynamic conditions. \medskip

Although these examples are simple, they demonstrate the flexibility of the ADT by extending it to model reaction processes. The compositional nature and modularity of the type system ensures that additional features not implemented here, such as catalysts or thermal energy, could be straightforwardly implemented. Stoichiometric or thermodynamic constraints could be enforced through use of dependent types, although this is left as future work. \medskip

\begin{center}
\begin{minipage}{\textwidth}
\begin{lstlisting}[caption={Haskell representation of a chemical reaction including reactants, products, conditions, and reaction rate. The \texttt{Reaction} data type is composed using the \texttt{Molecule} data type}, label={lst:reaction-representation}, language={Haskell}, captionpos=b]
-- Reactions or transformations between chemical species
data Reaction = Reaction
    { reactants :: [(Double, Molecule)]
    , products :: [(Double, Molecule)]
    , conditions :: [Condition]
    , rate :: Double
    }

-- Conditions under which a reaction occurs
data Condition = TempCondition {temperature :: Double}  
                | PressureCondition {pressure :: Double}

data Times = Times { startTime :: Double, endTime :: Double }
\end{lstlisting}
\end{minipage}

\begin{minipage}{\textwidth}
\begin{lstlisting}[caption={Example reaction: formation of water from hydrogen and oxygen},language={Haskell}, label={lst:reaction-example}, captionpos=b]
-- Example reaction: 2H2 + O2 -> 2H2O
exampleReaction :: Reaction
exampleReaction = Reaction
  { reactants = [(2.0, hydrogen), (1.0, oxygen)]
  , products = [(2.0, water)]
  , conditions =
      [ TempCondition 500.0
      , PressureCondition 1.0
      ]
  , rate = 0.1
  }
\end{lstlisting}
\end{minipage}
\end{center}
\medskip

In contrast, string-based representations such as reaction SMILES, SMARTS, and SMIRKS attempt to encode reactions as sequences of characters~\citep{david2020molecular}. Reaction SMILES, for example, separates reactants and products using the ``>>'' operator, but does not inherently encode reaction conditions, catalysts, or rate dependencies~\citep{wigh2022review}. SMARTS and SMIRKS provide pattern-matching capabilities for substructure searching and reaction templates, respectively, but lack a native mechanism for enforcing reaction validity~\citep{david2020molecular}. \medskip

String-based representations also inherit the limitations of the string data type itself. A lack of compositionality, and difficulties in formal reasoning make it unclear how one might extend SELFIES, for instance, to define a reaction SELFIES in which all syntactically valid strings correspond to chemically valid reactions~\citep{krenn2022selfies}. Even if such an extension were possible, enforcing network-wide properties such as reversibility, equilibrium conditions, or pathway constraints would be nontrivial. \medskip

To generalise what is written here to adequately capture the syntax of reactions and reaction networks, it would be useful to derive an implementation off of a categorical semantics such as \citep{baez2017compositional}. However, this is left as future work. Chemical reaction networks align naturally with category-theoretic formalisms, where reactions can be treated as morphisms between reactant and product sets~\citep{baez2017compositional}. This view has been used to model biochemical pathways, open dynamical systems, and Petri nets in systems chemistry~\citep{baez2017compositional},  making an ADT in a typed functional language with support for categorical constructs such as Functors. \medskip

While constraints such as  stoichiometric balance are not enforced in our implementation, the ADT provides a foundation upon which such constraints could be incorporated in future work through dependent types or additional verification layers. \medskip

However, the compositional nature of the ADT, its ability to be amended to incorporate reaction properties, and its alignment with category theoretic formalisms make it a natural framework for cheminformatics applications.

\subsection{Storage, Transmission, and Readability}
Molecules in the ADT can be serialised and stored as \texttt{.hs} files, which may be used for storage, and for transmission of molecular data, perhaps via large chemical databases, through straightforward use of the \texttt{Show} and \texttt{Read} instances in Haskell, as shown in Listing \ref{lst:read-write}. \medskip

\medskip
\noindent
\textbf{Parsing and construction.}
The SDF reader constructs the atom map from V2000 atom blocks (including coordinates and element attributes), populates \texttt{localBonds} from bond lines, applies formal charges from \texttt{M\ \ CHG} annotations, and detects common six‑membered alternating rings to build a single \(\pi\) system with \(s=6\) over the edges of the ring. Programmatic examples (e.g.\ benzene) mirror this layout six C–C \(\sigma\) edges, six C–H \(\sigma\) edges, and one labelled \(\pi\) system spanning the ring so text and code paths agree on \((V,B)\). 

\begin{minipage}{\textwidth}
\begin{lstlisting}[caption={Haskell code demonstrating the serialisation and writing to disk of a \texttt{Molecule} data type and the reading of a molecule stored in an \texttt{.hs} file, through use of the \texttt{show} and \texttt{read} functions.}, language={Haskell}, label={lst:read-write}, captionpos=b]
-- Writing the molecule to a file
writeMoleculeToFile :: FilePath -> Molecule -> IO ()
writeMoleculeToFile filePath molecule = writeFile filePath (show molecule)

-- Reading the molecule from a file
readMoleculeFromFile :: FilePath -> IO Molecule
readMoleculeFromFile filePath = do
  contents <- readFile filePath
  return (read contents)

-- Example usage
main :: IO ()
main = do
  -- Write the methane molecule to a file
  writeMoleculeToFile "methane.hs" methane
  
  -- Read the methane molecule from the file
  molecule <- readMoleculeFromFile "methane.hs"
  
  -- Print the molecule read from the file
  print molecule
\end{lstlisting}
\end{minipage}

\begin{definition}[Serialization]
\emph{Serialization is the process of converting an in-memory data structure into a standardized, external format (e.g.\ a byte stream or a textual encoding) so that it can be stored or transmitted and later reconstructed in its original form.}
\end{definition}

This code demonstrates the ease of serializing and deserializing molecules to and from files in Haskell. By leveraging the \texttt{Show} and \texttt{Read} instances defined for the Molecule data type and its related types, molecules can be seamlessly converted to and from their string representations, allowing them to be stored in files and retrieved later. There are algebraic laws governing these, e.g. show and then read must be the identity function and vice versa. \medskip

This functionality provides a convenient way to save and load molecular data, and is particularly useful when working with large molecules or when molecules are required across different program runs. Although storage and transmission may take place via serialisation into a human and machine-readable string of text, it is important to distinguish between the serialized format of the data and the underlying data type itself. While the ADT can be serialized into a text string, it is crucial to recognize that the ADT itself is not merely a string.

\subsection{Probabilistic Programming}

Probabilistic programming languages (PPLs) automate such inference on firm mathematical footing \citep{heunen2018semantic}; mature systems include Pyro \citep{bingham2018pyro}, Infer.NET \citep{InferNET18}, and Edward \citep{tran2018simple}, alongside modular, composable PPL frameworks \citep{goldstein:msc,scibior2015practical,nguyen2022modular}. Relative to conventional deep learning, Bayesian methods offer interpretability and principled treatment of multiple forms of uncertainty \citep{yang2017explainable,barber2012bayesian}, and have been applied to drug design (e.g., Bayesian neural networks for toxicity prediction \citep{semenova2020bayesian}). Bayesian inference inverts the usual programming direction: given observed outputs, infer latent inputs and parameters that most plausibly generated them. In cheminformatics, the inputs encode molecular features under a chosen representation; inference is meaningful only when \texttt{sample}/\texttt{score} are defined for that type (e.g., for a \texttt{String}, a prior and likelihood over strings must be specified). A generative PPL model exposes two primitives: \emph{sample} (draws from specified distributions, operationally reducible to base uniform noise) and \emph{score} (weights executions by the likelihood under observed data). Inference typically uses (i) Monte Carlo (e.g., MCMC), (ii) variational optimisation, or (iii) exact calculation when conjugacy permits. In this article we use Trace Metropolis Hastings \citep{heunen2018semantic}. For molecular representations, PPLs naturally encode structural uncertainty and property variability. \medskip

In \citep{perello2016background,flach2012machine}, there are discussions of the importance of classifiers providing not just a single output, but two key pieces of information: a class label and an associated probability. This dual output allows for a more nuanced understanding of the classifier's confidence in its predictions. For instance, in binary classification tasks, a classifier might assign a label (e.g., `spam' or `not spam') and also provide the probability of the instance belonging to that class. This probabilistic information is crucial for tasks such as ranking, where items are ordered based on the likelihood of belonging to a particular class, and for making informed decisions in applications where the costs of false positives and false negatives differ. \citep{flach2012machine} emphasizes that effective classification involves not only assigning the correct label but also accurately estimating the probability associated with that label to reflect the classifier's certainty. In the case of SMILES, an estimate of 0.5 that it the molecule is soluble may mean exactly that or it may be saying that it is an example that is has never seen before. Without a grammar that tells one what is \textbf{valid} and what is \textbf{invalid}, it relegates the probability estimates to be ambiguous and potentially meaningless. SMILES syntactic validity is decidable by parsing against the specification; the practical issue for generative modeling is that unconstrained decoders can emit invalid strings, requiring rejection sampling, constrained decoding, or alternative representations.

\begin{minipage}{\textwidth}
\begin{lstlisting}[language=Haskell,
  caption={Trace M-H sampler that grows a molecule and scores it via a linear logP model on molecular features (jitter=0.1, burn-in=1000) \citep{goldstein:msc}.},
  label={lst:probprog-combined}, captionpos=b, basicstyle=\footnotesize\ttfamily, numbers=none, columns=fullflexible]
-- Pseudo-Haskell; domain types & PPL ops assumed in scope; M = Data.Map.
type Params = (Double,Double,Double,Double,Double)

model :: Double -> Meas (Molecule, Params)
model obs = do
  let n = 3
  atoms <- forM [1..n] $ \i -> do
    s <- sample (uniformD [C,N,O,H])
    [x,y,z] <- replicateM 3 (sample (normal 0 1))
    pure Atom{ atomID=i, atomicAttr=elementAttributes s
             , coordinate=Coordinate x y z, shells=elementShells s }
  let pairs = [(i,j) | i <- [1..n], j <- [i+1..n]]
  bs <- fmap concat $ forM pairs $ \(i,j) -> do
          inc <- sample (uniformD [True,False])
          if not inc then pure [] else do
            k <- sample (uniformD [1,2,3])
            let b = Bond{ delocNum = 2*k, atomIDs = Nothing }
            pure [((i,j),b),((j,i),b)]
  let m = Molecule{ atoms=atoms, bonds=M.fromList bs }

  [b0,b1,b2,b3,b4] <- replicateM 5 (sample (normal 0 0.1))
  let s  = moleculeSize m
      w  = moleculeWeight m
      a  = moleculeSurfaceArea m
      bo = moleculeBondOrder m
      yhat = b0 + b1*s + b2*w + b3*a + b4*bo
  score (normalPdf obs 0.2 yhat)
  pure (m,(b0,b1,b2,b3,b4))

main :: IO ()
main = observedLogPIO >>= \obs ->
       mh 0.1 (model obs) >>= print . map fst . take 1000 . drop 1000
\end{lstlisting}
\end{minipage}

Some code is excluded from Listing~\ref{lst:probprog-combined} to conserve space, but its intent is described here. The module implements a single probabilistic program that \emph{grows} a candidate molecule and \emph{conditions} on a target property, the observed \(\log P\) (which is taken from \citep{classicalgsg_logp_zenodo}). The scalar observation is loaded by \texttt{observedLogPIO} (e.g., from an SDF/CSV record), and the model is run under a trace Metropolis--Hastings kernel with jitter \(0.1\), discarding the first \(1000\) iterations (burn-in) and retaining the next \(1000\) draws. \medskip

The molecular state uses a lightweight Algebraic Data Type. A \texttt{Molecule} is a record with fields \texttt{atoms :: [Atom]} and \texttt{bonds :: M.Map (Int,Int) Bond}. Each \texttt{Atom} carries \texttt{atomID :: Int}, \texttt{atomicAttr :: AtomAttr}, \texttt{coordinate :: Coordinate} (a 3D point), and \texttt{shells :: Shells}. A \texttt{Bond} stores at least \texttt{delocNum :: Int} (an order-derived delocalisation count) and optionally \texttt{atomIDs :: Maybe (Int,Int)}; in our representation the bond graph itself is keyed externally by atom pairs, so \texttt{atomIDs} may be \texttt{Nothing}. The map \texttt{bonds} is treated as symmetric by inserting both directions \((i,j)\) and \((j,i)\) for each undirected edge.\medskip

Within the generative function \texttt{model}, we first sample a small scaffold of \(n=3\) atoms. For each \(i \in \{1,\dots,n\}\), the element symbol is drawn uniformly from \texttt{[C,N,O,H]}; atomic attributes \texttt{elementAttributes} and \texttt{elementShells} are looked up from the symbol; and the Cartesian coordinates are drawn independently from \(\mathcal N(0,1)\) in each dimension. This produces \texttt{Atom\{atomID=i, atomicAttr, coordinate, shells\}} with contiguous identifiers \(1..n\). We then consider every unique unordered pair \((i,j)\) with \(i<j\) and flip a fair coin to decide whether a bond is present. If included, the bond order \(k\in\{1,2,3\}\) is sampled uniformly and mapped to \texttt{delocNum = 2k}; the resulting \texttt{Bond} is inserted under both keys \((i,j)\) and \((j,i)\) to enforce symmetry. The atoms and bonds are combined into a \texttt{Molecule} value \texttt{m}.\medskip

Property conditioning is applied through a linear predictor over standard molecular features. From \texttt{m} we compute \texttt{moleculeSize m}, \texttt{moleculeWeight m}, \texttt{moleculeSurfaceArea m}, and \texttt{moleculeBondOrder m}, denoted \(s,w,a,\) and \(bo\). The coefficients \(b_0,\dots,b_4\) have independent Gaussian priors \(b_i \sim \mathcal N(0,0.1^2)\). The predicted \(\log P\) is
\[
\hat y \;=\; b_0 + b_1 s + b_2 w + b_3 a + b_4\,bo,
\]
and the observed value \(y\) is given a Normal likelihood \(y \sim \mathcal N(\hat y,0.2^2)\). In a trace M--H implementation, 

a proposal perturbs a subset of the current random choices (e.g., atom labels, local $\sigma$-edges, and selected bonding-system parameters). For readability, the probabilistic-programming sketch uses a simplified subset of the full ADT; the full representation (including bonding systems) can be sampled analogously. The jitter hyperparameter controlling the proposal scale; acceptance is computed from the usual Metropolis ratio given the prior densities and the \(\log P\) likelihood.\medskip

This unified program couples structure generation with property supervision: the posterior concentrates mass on those atom/bond configurations that yield features consistent with the target \(\log P\), while simultaneously quantifying uncertainty in the regression parameters. Practically, this provides a compact vehicle for inverse design proposals that move the structure are immediately scored by the property model so the chain preferentially explores chemically plausible neighborhoods that explain the target. The same grammar-and-score paradigm transfers naturally to synthesis planning: generative moves propose intermediates, and probabilistic scores (learned or mechanistic) guide search, a pattern that aligns with successful retrosynthesis strategies in the literature \citep{retrosynthesis}. \medskip

\subsubsection{Why evaluation over SMILES is not sensical}

In a probabilistic programming setting, the generative model and inference procedure must define a total probability distribution with a well-defined score for every execution trace. For SMILES, this forces an explicit treatment of ``negative syntax'', i.e. what happens when a proposed string is not parseable as SMILES. If invalid strings are left outside the model’s domain (no likelihood defined), then standard samplers (MCMC, SMC, importance sampling) cannot correctly compute acceptance ratios or weights whenever proposals fall off-support. If, instead, invalid strings are assigned zero probability (e.g., by attempting to parse and applying a factor of $-\infty$, then na\"ve sampling over raw character sequences becomes practically unusable: proposals land in the invalid region overwhelmingly often, producing an inordinate amount of ``junk'', extreme rejection rates, and highly variable weights. Either way, ``sampling over SMILES strings'' without an explicit syntactic failure semantics is not a fair empirical evaluation of probabilistic inference—most compute is spent rediscovering the grammar rather than exploring the molecular distribution. Consequently, principled PPL formulations must either (i) restrict the support to valid SMILES by construction (grammar-/parser-constrained generators, typed domains), or (ii) incorporate explicit parse-failure handling in the model while using proposals that respect the syntactic manifold, so that inference efficiency and reported metrics reflect modeling quality rather than accidental syntax validity.

\subsection{Existing barriers to general molecular representations.}

Having introduced the MolADT design and its reference implementation, we now summarise the concrete representational requirements that most strongly affect ML evaluation protocols. These requirements motivate why we treat representation as a semantic contract rather than as a neutral serialisation, and they also define what should be reported (and ideally controlled) when benchmarking molecular ML models. String‑first encodings suffer from far more than localized bonding, hidden hydrogens, and patchy stereochemistry. They are \emph{syntactically} brittle (SMILES can be invalid) and, even when made syntactically robust (SELFIES), remain \emph{semantically} limited because they still decode to labeled 2D graphs \citep{krenn2020self,krenn2022selfies}. They are non‑unique and tool‑dependent (divergent canonicalisation and aromaticity perception), which injects noise into indexing and learning \citep{david2020molecular}. Their traversal‑based strings are non‑local with respect to chemistry, so string distances correlate poorly with geometric or energetic similarity \citep{fang2022geometry}. Critically, they omit 3D coordinates and torsions central for binding and stability, forcing external geometry generators and breaking round‑trips \citep{zhou2022uni,xu2022geodiff}. Binary edges with integer orders cannot express delocalisation, multi‑centre bonding, hapticity, or zero‑order interactions (e.g., diborane, ferrocene) without ad hoc labels \citep{wigh2022review,david2020molecular,dietz1995yet}. Coverage for non‑tetrahedral and axial/helical/planar stereochemistry is incomplete and inconsistent (e.g., cis/trans‑platin) and often needs 3D context to disambiguate \citep{wigh2022review,daylight_manual}. Tautomers/protomers and resonance are handled outside the representation; hydrogens are frequently implicit; ions, non‑covalent interactions, and spin are out of scope \citep{lipfert2014ionic}. Polymers, macromolecules, crystals, and periodicity fragment into bespoke grammars; reactions lack typed support for stoichiometry, conditions, and rates (pushed into side‑channels like RInChI/ProcAuxInfo) \citep{wigh2022review,doi:10.1021/acs.jctc.8b00640,heller2013inchi}. Overall, strings are convenient \emph{formats}, not principled \emph{data types}, and they misalign with Bayesian and geometric ML, which need explicit structure, 3D symmetry actions, and typed priors/likelihoods \citep{bronstein2021geometric,kendall2017uncertainties}. \medskip

Our ADT is designed to encode (and our reference implementation partially realises): Dietz-style bonding systems for delocalised/multicentre bonding; explicit atom-level annotations (e.g., charge, isotope, optional hydrogens); optional coordinates supporting stereo/conformer handling; and typed extension points for reaction objects and probabilistic modelling. Where features are sketched rather than fully implemented, we mark them explicitly as future work. \medskip

These limitations, the authors argue (and we agree), are an inherent problem with string-based representations in their ability to represent complex bonding (such as with diborane and ferrocene) and advanced stereochemical features. Cis- and trans-platin are also given as examples of molecules current string-based representations cannot represent. \medskip

The lack of quantum chemical information, which is not currently part of any modern digital representation, is also a barrier to representation, as expressed in \citep{krenn2020self}:  \medskip

``\textit{Thus, it should be stressed again that in d- and f-block chemistry, as well as main-group organometallic compounds, it is often impossible to assign any particular bond orders without high-level quantum chemical calculations, due to the highly delocalized nature of the bonding, where electrons are often spread out over a significant number of atoms, including the metal center itself, the immediately coordinated atoms, and additional ligand groups.}'' \medskip

Despite recognizing the need for a general digital molecular representation, capable of handling all of the above issues. The authors do not reconsider the suitability of string-based models, and instead, the solutions presented primarily involve introducing new grammars, expanding syntax rules, and further overloading symbols, leading to increasingly complex and specialized representations.

\textbf{``Molecular programming languages''.}\newline
In Future Project \#8, Krenn et al. (2022) propose a ``molecular programming language''. They reason that strings can encode the powerful computational algorithms: 
\begin{quote}
``Strings can store Turing-complete programming languages: In the most general case, one can store the source code of computer programs as strings. For example, a Python file is a `simple string', which is executed by the Python interpreter. Python is, of course, a Turing-complete language, which means that strings can encode the most powerful computational algorithms.''
\end{quote}\medskip

Although program source can be stored as text, treating a raw string as a \textbf{semantic} representation conflates storage with meaning: correctness properties live in parsers, type checkers, and interpreters, not in the character sequence itself. In the molecular setting, we therefore aim to make semantic structure explicit in the data model rather than relying on string-level mutation. \medskip

Furthermore, if the programming language is represented by a string, the question of how functions within the language are interpreted is unanswered. If they are also strings, they would at some point need to be executed by something other than a string. If they are not strings, then it remains to be understood why the authors claim that strings alone are enough to represent a Turing-Complete language, let alone the utility of such a project. \medskip

This line of work can be read as a desire for compositional syntax and semantics coupled with robust mutation operators. Our approach provides these by modelling molecules as typed values with explicit constructors and invariants; composition is then obtained using ordinary functions over the ADT. \medskip

Future Project \#9 goes beyond molecular programming languages, and proposes a programming language which is ``100\% robust'' by finding a syntax for a programming language that however you combine elements in the instruction set, a valid program is always produced. It seems this is an attempt to generalise their own robustness of mutation to the most general data structure: programs themselves. The feasibility and utility of ``robust programming-language'' objectives are outside our scope; we focus instead on ensuring that molecular representations have explicit constructors, well-scoped operations, and clear validity conditions. \medskip

We mention the definition of a domain specific language via a definitional interpreter \citep{10.1145/800194.805852}. This perspective allows the syntax of our \texttt{Molecule} data type to be embedded in a host language (shallow or deep), here Haskell, paired with an environment (a mapping from variables to values). In this work, we use this framing only as a conceptual bridge: rather than treating molecular strings as ``programs'', we emphasise molecules as typed values with well-scoped constructors and operations that support validation, transformation, and probabilistic modelling. \medskip

We also make progress on the future projects suggested:

\begin{itemize}

\item \textbf{Complicated bonds (Dietz \& zero-order bonds; context for FP6).} 
We adopt the Dietz-style constitution (multigraph of bonding systems) to capture multicenter bonding and encode ``zero-order'' interactions as bonds with 0 shared electrons in the ADT. 
Krenn \emph{et al.} note: ``Dietz suggested a hypergraph concept \dots accounting for multicenter bonding,'' and dashed interactions in diborane ``have been termed `zero-order bonds' by Clark'' \citep{krenn2022selfies}. 
\emph{This work:} represented as a bonding system with \texttt{sharedElectrons = 0} (a ``zero-order'' system) and an explicit member edge set; multicentre bonding is represented by bonding systems whose member edge set greater than one elements.

\item \textbf{FP6 — Generalization of SELFIES and automatic compilation of complex rules from data (partial).} 
Krenn \emph{et al.}: ``define a robust generalization of SELFIES that incorporates molecules beyond VBs'' \citep{krenn2022selfies}.  
\emph{This work:} we address the representational need (beyond valence-bond assumptions) \emph{not} by generalizing SELFIES, but by a typed constitution\,+\,3D configuration ADT (Dietz-style bonding systems, explicit H, advanced stereo, coordinates). We do \emph{not} attempt ``automatic compilation of complex rules from data.''

\item \textbf{FP2 — The effect of token overloading in generative models (design relevance only).} 
Krenn \emph{et al.}: ``One important question is to understand how overloading impacts ML models'' \citep{krenn2022selfies}.  
\emph{This work:} our ADT has \emph{no} overloaded tokens (typed constructors instead of strings), removing the issue by design; we do \emph{not} perform the proposed controlled study.

\item \textbf{FP7 — Graph-edit rules and metaSELFIES for reactions (related, not the proposed method).} 
Krenn \emph{et al.}: ``A syntactically robust reaction representation would most likely improve the performance \dots'' and should ``conserve the number of atoms \dots and the total charge'' \citep{krenn2022selfies}.  
\emph{This work:} we provide a typed \texttt{Reaction} ADT where such conservation checks are natural; we do \emph{not} implement metaSELFIES or a graph-edit rule DSL.

\end{itemize}
\subsection{Comparison to existing tools}\label{subsec:comparison}

Table~\ref{tab:comparison} situates our approach relative to widely used molecular representations and toolkits by focusing on the capabilities of the \emph{core representation}, rather than on downstream algorithms. String and identifier schemes such as SMILES/OpenSMILES, SELFIES, and InChI prioritise compactness, robustness, or canonicalisation, but they do not provide first-class support for three-dimensional structure, reactions, or general delocalised and multi-centre bonding; where aromaticity or resonance is handled, it is typically via flags or normalisation conventions rather than explicit structure. File formats such as SDF/Molfile can encode 2D/3D coordinates and stereochemistry, but function primarily as interchange containers rather than as editable, invariant-preserving internal models. Graph-based toolkits like RDKit and Open Babel offer rich chemistry operations, reaction handling, and broad interoperability, yet their internal representations still largely rely on conventional bond models with special-case treatments for delocalisation. In contrast, this work adopts a typed Algebraic Data Type as the primary molecular representation, enabling explicit, first-class treatment of general delocalised bonding alongside 3D geometry, stereochemistry, and reactions, while remaining open-source and suitable as a reference model rather than merely a file format or algorithmic library.

\begin{table}
\scriptsize
\setlength{\tabcolsep}{2.5pt}
\renewcommand{\arraystretch}{1.05}

\begin{tabularx}{\textwidth}{@{}l l c c c c c >{\raggedright\arraybackslash}X@{}}
\toprule
System &
Core rep. &
General deloc. &
3D &
Stereo &
Rxns &
OSI &
Notes \\
\midrule

\vtop{\hbox{\strut SMILES /}\hbox{\strut OpenSMILES}} &
String &
\no &
\no &
\yes &
\no &
-- &
Line notation for molecules; \\

SELFIES &
String &
\no &
\no &
\yes &
\no &
\yes &
Robust string representation (decoder guarantees validity under its grammar); \\

InChI &
Identifier &
\no &
\no &
\yes &
\no &
-- &
Canonical identifier for lookup and deduplication rather than editing; \\

SDF / Molfile &
File format &
\no &
\yes &
\yes &
\no &
-- &
Exchange format: can carry 2D/3D coordinates and stereo bond annotations; \\

RDKit &
Graph toolkit &
\no &
\yes &
\yes &
\yes &
\yes &
Mature cheminformatics toolkit with broad interop; \\

Open Babel &
Graph toolkit &
\no &
\yes &
\yes &
\yes &
\yes &
Broad format conversion and chemistry operations;\\

\textbf{This work} &
Typed ADT &
\yes &
\yes &
\yes &
\yes &
\yes &
Our reference implementation \\
\bottomrule
\end{tabularx}

\caption{High-level comparison of molecular representations and toolkits. Here
``General deloc.'' means \emph{first-class support for general delocalised/multi-centre
bonding} (beyond aromaticity flags / kekulisation conventions). Symbols:
\yes\ supported; \no\ not supported; ``--'' not applicable.}
\label{tab:comparison}
\end{table}

\subsubsection{Representation-aware benchmarking.}
When evaluating molecular ML models, it is rarely sufficient to report only dataset, architecture, and metric. The representation defines additional experimental conditions that should be held fixed—or at least reported—because they change the learning problem. At minimum, benchmarks should specify:

\begin{itemize}
    \item Validity contract: what structures are considered valid and how validity is enforced (parser-only, constrained decoding, rejection sampling, or representation-level invariants);
    \item Edit locality: what constitutes a ``small change'' (token edit, graph edit, or structured transformation) and how perturbations are generated;
    \item Symmetry handling: whether the model is required to be invariant/equivariant to atom-index permutations and (when 3D is present) rigid motions, and how this is enforced or tested.
\end{itemize}

MolADT is intended as a substrate that makes these explicit. By providing a typed semantic core with deterministic validation and well-scoped transformations, MolADT enables benchmarking studies that can ask representation-level questions directly for example, how model validity rates, novelty/diversity measures, uncertainty calibration, or sample efficiency change when the edit operations are local structure-preserving transformations rather than token mutations.

\subsection{Limitations and Future Work}
We see three concrete next steps.
\begin{enumerate}
\item \textbf{Coverage extensions.} Add explicit support for tautomer sets, ionic/non-covalent contacts, and polymer/repeat-unit abstractions.
\item \textbf{Interoperability.} Extend parsers and round-trip tests for SMILES/InChI and reaction formats (SMIRKS/RXN) so that comparisons can be made on shared corpora.
\item \textbf{Empirical evaluation.} Benchmark parsing/validation speed and memory, and assess downstream impact on a small set of standard tasks (e.g., property prediction with symmetry-aware models) where representation choice is known to matter.
\end{enumerate}

The statistical and computational efficiency of our reference implementation is explicitly not the focus of our article. Our current prototype prioritises clarity and correctness, over compactness. In practice, the ADT is intended as an in-memory representation; serialisation can target standard compact encodings (SDF/SMILES/InChI) or a dedicated binary format. We therefore treat file-size considerations as an engineering concern rather than a representational limitation. Furthermore, by storing a molecule as intensionally as a function or properties as higher-order functions, rather than extensionally, it may be possible to significantly reduce the storage size of a molecule.
\subsection{Future work}

The ADT so far presented has the ability to be improved upon, extended, and used in empirical experiments to verify and test the representation's utility, ease-of-use and verifiability across a number of domains in cheminformatics. Parsers from other formats e.g. SMILES strings to this ADT could also be used. Currently a parser for \texttt{.SDF} files has been implemented.

\subsubsection{Geometric Deep Learning on Molecular Representations}\label{gdl-on-reps}
Future work could consist of using symmetry-aware with Geometric Deep Learning, to reduce estimation error (requiring less data to infer parameters). A group captures the set of symmetrical operations over the representation. For molecules the key properties are (i) permutations of atom IDs and (ii) rigid 3D motions (rotations/translations). Randomized SMILES already tries to mimic (i) by enumerating alternative traversals; GDL would make this automatic any relabeling or rigid motion is handled identically (invariant for scalar properties, equivariant for vector fields). Practically, we can encode this contract in code by introducing a small Haskell Group typeclass, with instances for atom-index permutations and rigid motions; models then declare which actions they respect, and the type system enforces symmetry-safe use. This process of learning leverages algebraic structure in the domain, to reduce estimation, model and approximation error amongst other things.

\subsubsection{Extending and Conceptual Evaluation of the ADT}
\label{sec:adt-extensions}

The Algebraic Data Type presented here can be extended and empirically evaluated to test its utility across diverse tasks in cheminformatics. Two complementary lines of work are: (i) improving interoperability through additional parsers, and (ii) augmenting the core representation to cover chemical phenomena not yet encoded.

\paragraph{Interoperability and parsers.}
To broaden adoption, it is useful to support import from multiple established formats (e.g., SMILES) into the ADT; at present, we provide parsers only for SDF files. Extending the parsing layer will permit side-by-side evaluations and reduce friction when integrating the ADT into existing workflows.

\paragraph{Augmenting the representation}
To increase expressiveness, the ADT should be extended to represent tautomerism, ionic bonding, polymerism, and selected quantum information, such as spin, which are currently not captured.

\begin{description}
\item[\textbf{Tautomerism.}] Tautomers occur in many therapeutics, from sildenafil (Viagra) to warfarin, remdesivir (used for COVID-19), tetracyclines, and other antibiotics. Explicit handling of tautomeric states can expand the reachable chemical space and improve tasks such as ligand–protein binding prediction \citep{pajka2022theoretical,pospisil2003tautomerism,boyles2020learning}.

\item[\textbf{Ionic bonding.}] Ionic interactions are central to the stability and function of nucleic acids, proteins, and membranes, and to processes such as catalysis and ion transport \citep{lipfert2014ionic,baldwin1996hofmeister,petukh2014ion}. Encoding ions and their bonding explicitly would enable modeling of charged species and electrochemical properties, supporting the design of materials including ionic liquids and solid electrolytes for energy storage and green-chemistry applications \citep{macfarlane2017ionic}.

\item[\textbf{Polymerism.}] Many problems in chemistry and materials science involve polymers. By leveraging lists and recursive data structures, the ADT can be extended to represent polymeric and macromolecular architectures, enabling analysis and optimization of complex polymer systems.

\item[\textbf{Quantum information (spin).}] Incorporating basic quantum descriptors (e.g., spin multiplicity, localization of unpaired electrons) would allow the ADT to capture open-shell species and states relevant to reactivity and spectroscopy.
\end{description}

\subsubsection{Refinement Types via Liquid Haskell}

\textbf{Indeed, one can also use Liquid Haskell} \citep{vazou2016liquid} to encode simple forms of type-level constraints, known as \emph{refinement types}. A refinement type system allows us to refine existing types (e.g., \texttt{Int}, \texttt{Double}, etc.) with logical predicates, thereby adding extra \emph{compile-time} guarantees about program behavior. Unlike fully general dependent types, refinement types represent a compromise between expressive power and decidability. \medskip

By adding refinement types to Haskell, these constraints can be checked at \emph{compile time} rather than waiting until runtime (or even worse, detecting them only after a failed simulation). 

\subsubsection{Dependent Types}
Dependent types allow types to be parameterised by values, so domain invariants can be stated and checked by the type system at compile time. In our setting, they would permit encoding chemical constraints such as valence limits, charge balance, and valid connectivity directly in the type of a molecule, making many physically invalid structures unrepresentable \citep{norell2009dependently}. However, even with such expressive types, it remains unclear how one would ensure that the resulting probabilistic models employ chemically plausible priors or that their likelihood functions correspond to real, experimentally meaningful measurements.

\subsubsection{Development of a powerful, user-friendly library}
Future additions to the library can use Haskell's accelerate library \citep{10.1145/1926354.1926358} to parallelise computations from using monoidal structures (\emph{e.g.}, reductions, scans) being one common pattern for parallelism, \texttt{accelerate} supports a broad range of data-parallel operations on multidimensional arrays (maps, zips, permutations, stencils, etc.). In future engineering work, data-parallel backends (e.g., Accelerate) could be explored for accelerating specific kernels (e.g., traversal, validation) where benchmarks justify it. \citep{accelerate_hackage}). Functions should include the ability to convert between the ADT and other common representations, which would enable use of the ADT with existing potentially vast chemical databases. \medskip

Canonicalisation is also a concern for the ADT, as molecules may have different indices and may admit multiple coordinate realisations. The permutations of atom indices form a group, and these group properties can be leveraged when designing canonicalisation procedures and symmetry-aware comparisons. 

\subsubsection{Exploring the ADT Further}

To further strengthen the type safety and invariance guarantees of the representation, exploring dependently typed programming languages, such as Agda, could be highly beneficial. Dependent types allow for the encoding of more sophisticated invariants and constraints directly into the type system, ensuring that only valid and consistent molecular structures are expressible. This increased level of type safety can help catch potential errors at compile-time and provide stronger guarantees about the correctness of the representation. \medskip

Further work could explore several avenues to enhance the proposed Algebraic Data Type (ADT) representation and its applications. Integrating Haskell's Lens library \citep{steckermeier2015lenses} could simplify the manipulation of complex molecular structures by providing a composable, modular and type-safe way to access and modify nested data fields. \medskip

To enhance the modularity and composability of the ADT representation, exploring techniques like modular syntax trees and the ``Data Types a la Carte'' approach \citep{swierstra2008data} could prove highly valuable. By decomposing the molecular representation into smaller, reusable components, we can create a more flexible and adaptable framework. This modular design would allow for the easy integration of new features, the creation of domain-specific languages for molecular manipulation, and the development of reusable libraries and tools for cheminformatics. \medskip

These directions would clarify the practical scope of the representation and provide evidence for (or against) its utility on standard cheminformatics tasks, such as accelerating the discovery of novel and useful molecules. Implementing backpropagation using differentiable programming \citep{Elliott2009-beautiful-differentiation,vakar2022chad} would enable using gradient-based inference algorithms. \medskip 

We do not consider quantum-computing approaches here; our focus is the representational substrate and its validation properties. \medskip
 
At present, the efficiency and efficacy of the ADT in cheminformatics tasks are hypotheses. To validate the utility of the ADT as a molecular representation, empirical work is needed to benchmark its performance against representations such as SMILES and SELFIES, on tasks like molecular property prediction (e.g. calculation of logP and logS), virtual screening, and de novo drug design. \medskip

Another avenue for future work is using the proposed representation with geometric deep learning techniques to exploit symmetries and invariances in molecular structures \citep{bronstein2021geometric}. By leveraging mathematical properties of ADTs, such as group invariances and equivariances and their verification in Haskell through type classes, one could develop neural network architectures that are specifically tailored to the unique characteristics of those molecules. For instance, the choice of index for each \texttt{atomID} used in the representation should not influence a model, given the same molecule but with a different ordering of atomic IDs. \medskip

An additional avenue is using equational reasoning to investigate algebraic approaches in molecular modeling \citep{letychevskyi2021algebraic}. \medskip

Neural networks can be designed to be invariant to permutations of atomic indexes, ensuring that molecularly equivalent structures are treated identically regardless of labeling of indices or bonds. Similarly, molecular features such as the \texttt{atomSymbol}, are both translation and rotation invariant, while others may only be translation invariant. Incorporating these symmetries directly into the neural network architecture could significantly reduce the amount of training data required and improve the generalization capability of the models.

\section{Conclusion}
In this work, we introduced a molecular representation grounded in Algebraic Data Types and provided a reference implementation to make the method concrete and reproducible. The central idea is to treat molecules as structured, typed values rather than as strings so that validation, transformation, and composition are defined directly over the molecular structure and many malformed manipulations can be rejected early. The contribution is therefore not a claim of new state-of-the-art task performance, but a reference representation and implementation intended to enable more controlled, interpretable, and reproducible benchmarking studies in cheminformatics. \medskip

Our core constitution layer follows a Dietz-style valence multigraph formulation, which supports delocalised and multicentre bonding in a uniform way. We pair this with an 3D configuration layer (coordinates and stereochemical distinctions) and sketch how electronic annotations (shells, subshells, orbitals) can be carried when such metadata are available. We also show how the same typed approach naturally extends to reaction representations, where conservation checks and structural constraints can be expressed as ordinary program logic. \medskip

Overall, the contribution is a representation and reference implementation intended to support clearer reasoning, safer manipulation, and tighter integration with Bayesian and geometric machine-learning workflows than is typical with string-first formats. \medskip

Through Haskell's type classes and functional programming paradigm, we provide a rich and robust semantic context for molecular structure, and an efficient framework for computational tasks.  Molecular data adheres to the constraints of the molecular representation, going beyond the limitations of strings as a representation. \medskip

To demonstrate the ADT and how it can be extended, we explored group based properties of molecular symmetry. We also showed integration of the ADT with probabilistic programming, showing that the ADT can be appropriately used for Bayesian inference, and that machine learning processes can operate directly on the program or grammar itself, without the need for external context or to encode the representation to other formats. \medskip

The ADT presented was not intended to be a definitive representation, but a valuable representational concept for cheminformaticians. Nevertheless, the ADT provides a structured alternative to string-first formats (e.g., SMILES/SELFIES) for representing, transforming, and validating molecular structures, particularly in workflows that benefit from explicit invariants and compositional operations.

\section{Declarations}

\subsection{Availability and requirements}
\begin{description}
  \setlength\itemsep{0.2em}%
  \item[Project name:] Project name: MolADT-Bayes (reference implementation)
  \item[Project home page:] Project home page: \citep{new_repo}
  \item[Archived version:] \href{https://doi.org/10.5281/zenodo.18238032}{doi.org/10.5281/zenodo.18238032} (v1.0.4)
  \item[Operating system(s):] Tested on macOS Tahoe 26.2 (GHC 9.6.5, Cabal 3.14.2.0, Stack 3.7.1, ghcup 0.1.50.2)
  \item[Programming language:] Haskell (GHC; Stack or Cabal)
  \item[Other requirements:] GHC via Stack; build with \texttt{stack build}; run \texttt{stack exec moladtbayes}. A step-by-step \texttt{README.md} on the GitHub documents OS, dependencies, and exact commands.
  \item[License:] AGPL-3.0 (OSI-approved)
  \item[Restrictions to use by non-academics:] None beyond AGPL-3.0 obligations
\end{description}

The DB1/DB2 naming convention on the GitHub follows the dataset definitions introduced by Donyapour et al. in their ClassicalGSG study of the SAMPL7 logP challenge \citep{donyapour2021classicalgsg}. In that work, the authors construct a master training corpus (denoted DB1) by aggregating several publicly available experimental logP datasets (summarised in their Table 1). A chemically restricted subset, DB2, is then obtained by filtering DB1 to molecules containing only the elements C, N, O, S, and H, in order to match the elemental composition of the SAMPL7 target compounds. The same DB1/DB2 split is mirrored in our distribution (MolADT-Bayes/logp/DB1.sdf and MolADT-Bayes/logp/DB2.sdf) to enable direct comparability with prior results. These datasets, together with multiple predefined training/test splits and associated molecular and property files, are publicly available as the ``ClassicalGSG logP dataset'' via Zenodo \citep{classicalgsg_logp_zenodo}.

\subsection{Competing interests}
The authors declare no competing interests. The repository is listed on an Open Source Molecular Modeling Index on GitHub \citep{osmm_index}.

\subsection{Funding}
Oliver Goldstein was funded in part by an EPSRC DTP Scholarship and is also partly self-funded. Sam March is self-funded.

\subsection{Authors' contributions}

\subsubsection{Oliver Goldstein}
Ideated the project and conceptualized the solution in the form of an Algebraic Data Type, implemented the reference prototype and the worked examples/experiments, and drafted the initial article.
\subsubsection{Samuel March}
Made substantial revisions to the article's form and content, contributed to the research, and made miscellaneous contributions to the library.

\subsection{Acknowledgments}
We are thankful for discussions at the Algebra of Programming group at Oxford University, and comments by Tom Smeding, Sean Moss and Swaraj Dash.

\section{Appendices}
\begin{appendices}

\setcounter{lstlisting}{0}
\renewcommand{\thelstlisting}{A.\arabic{lstlisting}}

\begin{lstlisting}[
  label={lst:benzene-adt},
  language=Haskell,
  caption={Benzene with one Dietz $\pi$ pool ($s=6$) over the ring edges plus $\sigma$ edges; coordinates in \AA. Source : \citep{pubchem_benzene_cid241}.},
  frame=single,
  basicstyle=\ttfamily\small,
  columns=fullflexible
]
benzenePretty :: Molecule
benzenePretty = Molecule
  { atoms      = atomTable
  , localBonds = sigmaFramework
  , systems    = [(SystemId 1, piRingSystem)]
  }
  where
    -- Atom IDs
    carbons   = AtomId <$> [1..6]
    hydrogens = AtomId <$> [7..12]

    -- Shared element data (kept shared, like your original)
    carbonAttributes   = elementAttributes C
    hydrogenAttributes = elementAttributes H
    carbonShells       = elementShells C
    hydrogenShells     = elementShells H

    -- Small helpers
    coord (x,y,z) =
      Coordinate (mkAngstrom x) (mkAngstrom y) (mkAngstrom z)

    mkAtom aid attrs sh xyz = Atom
      { atomID       = aid
      , attributes   = attrs
      , coordinate   = coord xyz
      , shells       = sh
      , formalCharge = 0
      }

    mkEdges = S.fromList . map (uncurry mkEdge)

    -- Geometry (same numbers, just data-driven)
    carbonCoords =
      [ (-1.2131, -0.6884, 0.0)
      , (-1.2028,  0.7064, 0.0)
      , (-0.0103, -1.3948, 0.0)
      , ( 0.0104,  1.3948, 0.0)
      , ( 1.2028, -0.7063, 0.0)
      , ( 1.2131,  0.6884, 0.0)
      ]

    hydrogenCoords =
      [ (-2.1577, -1.2244, 0.0)
      , (-2.1393,  1.2564, 0.0)
      , (-0.0184, -2.4809, 0.0)
      , ( 0.0184,  2.4808, 0.0)
      , ( 2.1394, -1.2563, 0.0)
      , ( 2.1577,  1.2245, 0.0)
      ]

    carbonAtoms =
      zipWith (\aid xyz -> mkAtom aid carbonAttributes carbonShells xyz)
              carbons
              carbonCoords

    hydrogenAtoms =
      zipWith (\aid xyz -> mkAtom aid hydrogenAttributes hydrogenShells xyz)
              hydrogens
              hydrogenCoords

    allAtoms = carbonAtoms ++ hydrogenAtoms

    atomTable = M.fromList [(atomID a, a) | a <- allAtoms]

    -- Bonds
    ringPairs  = zip carbons (tail (cycle carbons))  -- (c1,c2) ... (c6,c1)
    chPairs    = zip carbons hydrogens               -- (c1,h7) ... (c6,h12)

    sigmaFramework = mkEdges (ringPairs ++ chPairs)
    piRingEdges    = mkEdges ringPairs

    piRingSystem = mkBondingSystem (NonNegative 6) piRingEdges (Just "pi_ring")
\end{lstlisting}

\begin{lstlisting}[
  label={lst:diborane-adt},
  language=Haskell,
  caption={Diborane (B2H6) with two 3c–2e bridges as Dietz pools.},
  frame=single,
  basicstyle=\ttfamily\small,
  columns=fullflexible
]
module Ferrocene (ferrocenePretty) where

import qualified Data.Map.Strict as M
import qualified Data.Set as S

import Chem.Dietz
  ( AtomId(..)
  , SystemId(..)
  , NonNegative(..)
  , mkEdge
  , mkBondingSystem
  )
import Chem.Molecule
  ( AtomicSymbol(..)
  , Molecule(..)
  , Atom(..)
  , Coordinate(..)
  , mkAngstrom
  )
import Constants (elementAttributes, elementShells)

ferrocenePretty :: Molecule
ferrocenePretty = Molecule
  { atoms      = atomTable
  , localBonds = sigmaFramework
  , systems    =
      [ (SystemId 1, cp1PiSystem)
      , (SystemId 2, cp2PiSystem)
      , (SystemId 3, feBackDonationSystem)
      ]
  }
  where
    -- Atom IDs
    fe      = AtomId 1
    ring1C  = AtomId <$> [2..6]
    ring2C  = AtomId <$> [7..11]
    ring1H  = AtomId <$> [12..16]
    ring2H  = AtomId <$> [17..21]

    -- Shared element data
    feAttributes       = elementAttributes Fe
    carbonAttributes   = elementAttributes C
    hydrogenAttributes = elementAttributes H

    feShells       = elementShells Fe
    carbonShells   = elementShells C
    hydrogenShells = elementShells H

    -- Helpers
    coord (x,y,z) =
      Coordinate (mkAngstrom x) (mkAngstrom y) (mkAngstrom z)

    mkAtom aid attrs sh xyz = Atom
      { atomID       = aid
      , attributes   = attrs
      , coordinate   = coord xyz
      , shells       = sh
      , formalCharge = 0
      }

    mkEdges = S.fromList . map (uncurry mkEdge)

    ringPairs xs = zip xs (tail (cycle xs))

    -- (Replace with PubChem SDF coords if desired.)
    feCoord = (0.0000,  0.0000,  0.0000)

    ring1CarbonCoords =
      [ ( 1.1800,  0.0000,  1.6600)
      , ( 0.3647,  1.1220,  1.6600)
      , (-0.9547,  0.6935,  1.6600)
      , (-0.9547, -0.6935,  1.6600)
      , ( 0.3647, -1.1220,  1.6600)
      ]

    ring2CarbonCoords =
      [ ( 0.9547,  0.6935, -1.6600)
      , (-0.3647,  1.1220, -1.6600)
      , (-1.1800,  0.0000, -1.6600)
      , (-0.3647, -1.1220, -1.6600)
      , ( 0.9547, -0.6935, -1.6600)
      ]

    ring1HydrogenCoords =
      [ ( 2.2700,  0.0000,  1.6600)
      , ( 0.7016,  2.1582,  1.6600)
      , (-1.8364,  1.3338,  1.6600)
      , (-1.8364, -1.3338,  1.6600)
      , ( 0.7016, -2.1582,  1.6600)
      ]

    ring2HydrogenCoords =
      [ ( 1.8364,  1.3338, -1.6600)
      , (-0.7016,  2.1582, -1.6600)
      , (-2.2700,  0.0000, -1.6600)
      , (-0.7016, -2.1582, -1.6600)
      , ( 1.8364, -1.3338, -1.6600)
      ]

    feAtom = mkAtom fe feAttributes feShells feCoord

    ring1CarbonAtoms =
      zipWith (\aid xyz -> mkAtom aid carbonAttributes carbonShells xyz)
              ring1C ring1CarbonCoords

    ring2CarbonAtoms =
      zipWith (\aid xyz -> mkAtom aid carbonAttributes carbonShells xyz)
              ring2C ring2CarbonCoords

    ring1HydrogenAtoms =
      zipWith (\aid xyz -> mkAtom aid hydrogenAttributes hydrogenShells xyz)
              ring1H ring1HydrogenCoords

    ring2HydrogenAtoms =
      zipWith (\aid xyz -> mkAtom aid hydrogenAttributes hydrogenShells xyz)
              ring2H ring2HydrogenCoords

    allAtoms = feAtom : (ring1CarbonAtoms ++ ring2CarbonAtoms ++ ring1HydrogenAtoms ++ ring2HydrogenAtoms)

    atomTable = M.fromList [(atomID a, a) | a <- allAtoms]

    -- adjacency (localised bonds): C–C rings + C–H
    ring1CCPairs = ringPairs ring1C
    ring2CCPairs = ringPairs ring2C
    ring1CHPairs = zip ring1C ring1H
    ring2CHPairs = zip ring2C ring2H

    sigmaFramework =
      mkEdges (ring1CCPairs ++ ring2CCPairs ++ ring1CHPairs ++ ring2CHPairs)

    -- Dietz-style bonding systems (electron pools)
    feToRing1 = [(fe, c) | c <- ring1C]
    feToRing2 = [(fe, c) | c <- ring2C]
    feToAll   = feToRing1 ++ feToRing2

    cp1Edges = mkEdges (feToRing1 ++ ring1CCPairs)
    cp2Edges = mkEdges (feToRing2 ++ ring2CCPairs)
    feBackEdges = mkEdges feToAll

    cp1PiSystem =
      mkBondingSystem (NonNegative 6) cp1Edges (Just "cp1_pi")

    cp2PiSystem =
      mkBondingSystem (NonNegative 6) cp2Edges (Just "cp2_pi")

    feBackDonationSystem =
      mkBondingSystem (NonNegative 6) feBackEdges (Just "fe_backdonation")
\end{lstlisting}

\begin{lstlisting}[
  label={lst:ferrocene-adt},
  language=Haskell,
  caption={Dietz-style bonding systems (paper): localized C–H and C–C bonds in localBonds ($\sigma$ adjacency), 6e pool over (Fe–C + ring C–C) for each Cp ring, 6e pool over all Fe–C edges},
  frame=single,
  basicstyle=\ttfamily\small,
  columns=fullflexible
]
module Diborane (diboranePretty) where

import qualified Data.Map.Strict as M
import qualified Data.Set as S

import Chem.Dietz
  ( AtomId(..)
  , SystemId(..)
  , NonNegative(..)
  , mkEdge
  , mkBondingSystem
  )
import Chem.Molecule
  ( AtomicSymbol(..)
  , Molecule(..)
  , Atom(..)
  , Coordinate(..)
  , mkAngstrom
  )
import Constants (elementAttributes, elementShells)

diboranePretty :: Molecule
diboranePretty = Molecule
  { atoms      = atomTable
  , localBonds = sigmaFramework
  , systems    =
      [ (SystemId 1, bridgeH3System)
      , (SystemId 2, bridgeH4System)
      ]
  }
  where
    -- Atom IDs
    b1 = AtomId 1
    b2 = AtomId 2

    h3 = AtomId 3  -- bridge
    h4 = AtomId 4  -- bridge

    h5 = AtomId 5  -- terminal on b1
    h6 = AtomId 6  -- terminal on b1
    h7 = AtomId 7  -- terminal on b2
    h8 = AtomId 8  -- terminal on b2

    -- Shared element data
    boronAttributes    = elementAttributes B
    hydrogenAttributes = elementAttributes H

    boronShells    = elementShells B
    hydrogenShells = elementShells H

    -- Helpers
    coord (x,y,z) =
      Coordinate (mkAngstrom x) (mkAngstrom y) (mkAngstrom z)

    mkAtom aid attrs sh xyz = Atom
      { atomID       = aid
      , attributes   = attrs
      , coordinate   = coord xyz
      , shells       = sh
      , formalCharge = 0
      }

    mkEdges = S.fromList . map (uncurry mkEdge)

    -- Idealised D2h-like geometry, chosen to make bridges explicit.
    bCoords =
      [ (-0.8850,  0.0000,  0.0000)  -- B1
      , ( 0.8850,  0.0000,  0.0000)  -- B2
      ]

    hCoords =
      [ ( 0.0000,  0.0000,  0.9928)  -- H3 bridge
      , ( 0.0000,  0.0000, -0.9928)  -- H4 bridge
      , (-0.8850,  1.1900,  0.0000)  -- H5 terminal (B1)
      , (-0.8850, -1.1900,  0.0000)  -- H6 terminal (B1)
      , ( 0.8850,  1.1900,  0.0000)  -- H7 terminal (B2)
      , ( 0.8850, -1.1900,  0.0000)  -- H8 terminal (B2)
      ]

    boronAtoms =
      zipWith (\aid xyz -> mkAtom aid boronAttributes boronShells xyz)
              [b1,b2]
              bCoords

    hydrogenAtoms =
      zipWith (\aid xyz -> mkAtom aid hydrogenAttributes hydrogenShells xyz)
              [h3,h4,h5,h6,h7,h8]
              hCoords

    allAtoms = boronAtoms ++ hydrogenAtoms
    atomTable = M.fromList [(atomID a, a) | a <- allAtoms]

    -- adjacency: B–B and four terminal B–H bonds
    sigmaFramework =
      mkEdges [ (b1,b2)
              , (b1,h5), (b1,h6)
              , (b2,h7), (b2,h8)
              ]

    -- 3c–2e bridges as Dietz pools (2 electrons shared over two edges)
    bridgeH3Edges = mkEdges [(b1,h3), (b2,h3)]
    bridgeH4Edges = mkEdges [(b1,h4), (b2,h4)]

    bridgeH3System =
      mkBondingSystem (NonNegative 2) bridgeH3Edges (Just "bridge_h3_3c2e")

    bridgeH4System =
      mkBondingSystem (NonNegative 2) bridgeH4Edges (Just "bridge_h4_3c2e")
\end{lstlisting}

\end{appendices}


\bibliography{sn-bibliography}

\end{document}